\documentclass[preprint2]{aastex631}

\usepackage{amsmath}
\usepackage{bm}
\usepackage{graphicx}

\begin{document}

% \title{Physics motivated neutron star hotspot map computation}
\title{Physics motivated models of pulsar X-ray hotspots: off-center dipole configurations}

\author[0000-0001-6406-1003]{Chun Huang}
\affil{Physics Department and McDonnell Center for the Space Sciences, Washington University in St. Louis; MO, 63130, USA}

\author[0000-0002-4738-1168]{Alexander Y. Chen}
\affil{Physics Department and McDonnell Center for the Space Sciences, Washington University in St. Louis; MO, 63130, USA}

\correspondingauthor{Chun Huang}
\email{chun.h@wustl.edu}


\begin{abstract}
    Recently, it was proposed that an off-center dipole magnetic configuration, together with a non-trivial temperature profile, may be the best model to explain the X-ray light curve of PSR J0030+0451 observed by the Neutron Star Interior Composition Explorer (\emph{NICER}). Using a theoretical model for the electric current density in a force-free pulsar magnetosphere, we compute from first principles the distribution of electric current over the polar cap associated with an off-center magnetic dipole. We then use a simple prescription to compute the resulting temperature distribution, which allows us to derive the observed X-ray light curve. We investigate the role of the volumetric return current region in the polar cap and find that although it does not make a big difference in an aligned dipole case, the difference can be bigger in the case of an off-center dipole. Finally, we apply Markov Chain Monte Carlo (MCMC) fitting to the X-ray light curves of pulsars PSR J0030+0451 and PSR J0437--4715 with and without the volumetric return current, and find that our model can reasonably recover the observed X-ray light curves.
\end{abstract}
\section{Introduction}
X-ray pulse profiles modeling constitutes a crucial methodology for
deducing key parameters of neutron stars such as their mass and radius. Typically, this type of modeling involves reverse-engineering observed X-ray light curves and spectra to infer the temperature distribution on the stellar surface, e.g.\ the locations and temperatures of hotspots~\citep[see e.g.][]{Morsink2007,B2013,2014ApJ...792...87P,Vincent2018}.

With the advent of the Neutron star Interior Composition Explorer (\emph{NICER})
and its high-quality timing data, precision modeling of the X-ray pulse profiles
of millisecond pulsars became a possibility. \citet{Miller2019} and
\citet{Riley2019} made the first measurement of the mass and radius of PSR
J0030+0451 by applying detailed modeling techniques to
\emph{NICER} data. The mass-radius measurement holds substantial implications,
providing a means to constrain the equation of state for matter inside the neutron star,
probing nuclear physics at these extreme conditions that are not accessible on
Earth~\citep[See e.g.][]{Raaijmakers_2019,Raaijmakers2020,Raaijmakers2021,Huang2024,Huang:2024rvj,Rutherford2024}.

In addition to determining the mass and radius of PSR J0030+0451,
\citet{Riley2019,Miller2019}
produced, at the time, the most detailed hotspot map of the neutron star surface derived
from observational data.
The hotspots were found to have distinct shapes: one is approximately circular, while the other is like a
  crescent. Both of the hotspots were found to be located within the same
hemisphere, suggesting magnetic structures beyond the regular dipole close
  to the star~\citep{Bilous_2019}. A few theoretical works followed, attempting to study the implication of the hotspot shapes on the magnetic field structure near the star~\citep{Chen_2020,Kalapotharakos2021}, and finding that a dipole plus quadrupole magnetic field can typically reproduce the observed X-ray light curve.

 A recent investigation by
\citet{Vinciguerra2024} revisited
the hotspot modeling of PSR J0030+0451, finding the hotspots to be consistent with the polar caps of an off-center dipole magnetic field but with a
nontrivial temperature distribution due to the implementation of new background constraints. In particular, the new study found that one of the hotspots has a high-temperature component near its edge. This study suggests that nontrivial temperature distribution across hotspots may need to be taken into account during the modeling process, which may significantly expand the parameter space. A theory that describes the temperature distribution across pulsar surface hotspots is urgently needed.

Fortunately, recent advances in theoretical modeling of the pulsar magnetosphere have provided a basis for such a theory. \citet{Gralla2017} have outlined a model to compute the magnetospheric current distribution across the neutron star surface, and \citet{Lockhart2019} have applied this model to compute potential surface temperature maps given a dipole plus axisymmetric quadrupole magnetic field configuration. However, the analytic models based on magnetospheric current have been limited to setups that are symmetric with respect to some magnetic axis. None of the hotspot configurations from phenomenological X-ray light curve modeling~\citep[e.g.][]{Miller2019,Riley2019} are axisymmetric. Going beyond axisymmetric models is a crucial step to take in order to bridge the gap between our theoretical understanding of pulsar magnetospheres and the observed X-ray data.

In this paper, we attempt to take a first step towards a non-axisymmetric, physics-based model of the temperature distribution across hotspots on the surfaces of rotation-powered pulsars. Based on the theory of force-free pulsar magnetospheres developed by \cite{Gralla2017}, we consider an off-center magnetic dipole and derive the distribution of the current density across the neutron star surface.  Using this current distribution, we compute the surface heating rate using a simple prescription and derive the surface temperature distribution. Finally, we compute the pulsar’s light curve using the \emph{X-PSI} global temperature module and incorporate it into pulse-profile modeling process. We perform a direct fit to the observed X-ray light-curve data of PSR~J0030+0451 (J0030) \citep{Bogdanov_2019b} and PSR~J0437--4715 (J0437) \citep{Choudhury_2024}, aiming to showcase the explanatory power of our off-center dipole temperature map. 
This work illustrates the potential of adopting a physics-based hotspot configuration to model pulsar X-ray data, offering an alternative to the geometric methods currently in use \citep[e.g., those proposed by][]{Riley2019}.

This paper is organized as follows. In Section~\ref{sec:method}, we provide a detailed description of the theoretical method. In Section~\ref{sec:results}, we compare the surface current distribution for shifted dipole models with results from canonical centered dipole models, highlighting the differences in hotspot structure and resulting X-ray light curves. In Section~\ref{sec:fitting}, we apply this model to try to fit the light curves of millisecond pulsars J0030 and J0437. Finally in Section~\ref{sec:discussion} we discuss the limitations of this method and outline future directions for this line of work.

\section{Method}
\label{sec:method}

We follow the general method developed by~\citet{Gralla2017} and \citet{Lockhart2019} to compute the current distribution on the surface of the pulsar. In this section, we outline the key ingredients, while referring the reader to the original paper for more detail.

The pulsar magnetosphere is assumed to be well-described by force-free electrodynamics~\citep[see e.g.][]{1999astro.ph..2288G,1999ApJ...511..351C}. Under such an assumption, the electric current flowing in the magnetosphere is entirely determined by the magnetic field configuration. \citet{2014MNRAS.445.2500G} showed that in this limit the electromagnetic field can be described using only two scalar fields $\alpha$ and $\beta$, which are often called ``Euler potentials''~\citep{1997PhRvE..56.2181U}:
\begin{equation}
    \label{eq:euler-potentials}
    F = d\alpha \wedge d\beta.
\end{equation}
Magnetic field lines are given by the intersection of surfaces of constant $\alpha$ and $\beta$. The Euler potentials satisfy the simple geometric equation:
\begin{equation}
    \label{eq:ffe-equations}
    \alpha\wedge d * F = \beta\wedge d * F = 0.
\end{equation}
Under axisymmetry, one of the Euler potentials becomes trivial, $\beta = \phi$, and the other equation can be reduced to the well-known Grad-Shafranov equation or pulsar equation~\citep[see e.g.][]{1973ApJ...180..207M,1999ApJ...511..351C} Solving the full non-axisymmetric problem in full 3D is still a challenging problem and often done using time-dependent numerical simulations~\citep[e.g.][]{2006ApJ...648L..51S}.

If a solution is already found, \citet{Gralla2017} showed that the four-current density in the pulsar magnetosphere can be computed directly from the Euler potentials:
\begin{align}
J^{\hat{t}}&=\sqrt{1-\frac{2 M}{r}} J^t, \label{eq:jt-hat}\\
J^{\hat{r}}&=\frac{\Lambda(\alpha, \beta)\left(\partial_\theta \alpha \partial_\phi \beta-\partial_\phi \alpha \partial_\theta \beta\right)}{\sqrt{r(r-2 M)}(r \sin \theta)}, \label{eq:jr-hat}\\
J^{\hat{\theta}}&=\frac{\Lambda(\alpha, \beta)}{r \sin \theta}\left(\partial_\phi \alpha \partial_r \beta-\partial_r \alpha \partial_\phi \beta\right), \label{eq:jth-hat}\\
J^{\hat{\phi}}&=\frac{\Lambda(\alpha, \beta)}{r}\left(\partial_r \alpha \partial_\theta \beta-\partial_\theta \alpha \partial_r \beta\right),\label{eq:jph-hat}
\end{align}
Where $(J^{\hat{t}}, J^{\hat{r}}, J^{\hat{\theta}}, J^{\hat{\phi}})$ are the components of the four-current components under the orthonormal basis $(\hat{\bm{t}}, \hat{\bm{r}}, \hat{\bm{\theta}}, \hat{\bm{\phi}})$, and $J^{t}$ is given by:
\begin{equation}
\begin{split}
J^t= & \frac{\Omega-\Omega_Z}{r(r-2 M)}\left\{\partial_\theta \alpha \partial_\theta \partial_\phi \beta-\partial_\theta \beta \partial_\theta \partial_\phi \alpha\right. \\
& +r(r-2 M)\left(\partial_r \alpha \partial_r \partial_\phi \beta-\partial_r \beta \partial_r \partial_\phi \alpha\right) \\
& -\partial_\phi \alpha\left[\left(1-\frac{2 M}{r}\right) \partial_r\left(r^2 \partial_r \beta\right)\right.\\
&\left.+\frac{\partial_\theta\left(\sin \theta \partial_\theta \beta\right)}{\sin \theta}\right] +\partial_\phi \beta\left[\left(1-\frac{2 M}{r}\right) \right.\\
&\left.\partial_r\left(r^2 \partial_r \alpha\right)
+\frac{\partial_\theta\left(\sin \theta \partial_\theta \alpha\right)}{\sin \theta}\right]\},
\end{split}
\label{eq:jt}
\end{equation}
where $\Omega$ is the angular velocity, $\Omega_Z$ is the ``frame-drag frequency'' defined as $\Omega_Z=2 \hat{I}\Omega/r^3 $, and $\hat{I}$ is the moment of inertia, which defined as $\hat{I} = \mathcal{I}M_{\star}R_{\star}^2$, where we set the compactness $\mathcal{I} = 2/5$. $M_{\star}$ and $R_{\star}$ are the mass and radius of the neutron star respectively. {The normalized time component $J^{\hat{t}}$ is the force-free charge density, and can be identified with the Goldreich-Julian charge density $\rho_\mathrm{GJ}$ first defined by~\citet{1969ApJ...157..869G}.}
% {Because $J^{t}$ represents the electric-charge density measured by a static observer, we identify it with the Goldreich–Julian (GJ) charge density, $ \rho_{\mathrm{GJ}}\;\equiv\;J^{t}. $ Throughout what follows we use this identification when forming the dimensionless ratio $j/(c\,\rho_{\mathrm{GJ}})$ that controls pair formation and polar-cap heating}.
The function $\Lambda(\alpha, \beta)$ is a conserved quantity along each field line that determines the magnitude of the current density. Comparing with first-principles PIC simulations of the global pulsar magnetosphere, \citet{Gralla2017} were able to find an approximate expression for this function:
\begin{equation}
\begin{aligned}
\Lambda(\alpha, \beta)= & \mp 2 \Omega\left\{J_0\left(2 \arcsin \sqrt{\alpha / \alpha_o}\right) \cos \iota\right. \\
& \left.\mp J_1\left(2 \arcsin \sqrt{\alpha / \alpha_o}\right) \cos \beta \sin \iota\right\}, \\
& \alpha<\alpha_o,
\end{aligned}
\label{lambda}
\end{equation}
where the $\mp$ sign refers to the northern/southern hemisphere, $\iota$ is the dipole inclination angle, $J_{0}$ and $J_{1}$ are
Bessel functions of the first kind, and $\alpha_0$ label the last open field line, defined as:
\begin{equation}
\label{eq:alpha0}
\alpha_o=\sqrt{\frac{3}{2}} \mu \Omega\left(1+\frac{1}{5} \sin ^2 \iota\right).
\end{equation}
This quantity $\alpha_{0}$ delineates the boundary of the polar cap, which is the region where open magnetic field lines originate from. The dipole magnetic moment $\mu$ is defined as $B_0\, R_{\star}^{3}$, where $B_0$ represents the magnetic field strength on the surface of the neutron star. In the subsequent sections, we adopt a value of $B_0 = 5 \times 10^8\,\mathrm{G}$ based on the rough spin-down estimation for millisecond pulsar populations. Equation~\eqref{eq:alpha0} is our \emph{definition} of the pulsar polar cap for the rest of this paper, even when the designated regions are no longer geometrically similar to polar caps in the traditional sense.

For a general rotating neutron star, no known analytic solution exists for the force-free Euler potentials $\alpha$ and $\beta$.
However, if we only consider a non-relativistic dipole-like magnetic field originating from the star, then a simple analytic solution exists for the Euler potentials:
\begin{equation}
\alpha =-\frac{\mu}{r}\sin ^2 \theta^{\prime}, \quad \beta =\varphi^{\prime},
\label{eq:euler-dipole}
\end{equation}
where $\theta^{\prime}$ and $\varphi^{\prime}$ are the polar coordinates about the dipole axis, and they are related to the coordinates $\theta$ and $\varphi$ in the lab frame by:
\begin{align}
\cos \theta^{\prime} & =\cos \theta \cos \iota-\sin \theta \cos \varphi \sin \iota \\
\tan \varphi^{\prime} & =\frac{\sin \theta \sin \varphi}{\sin \theta \cos \varphi \cos \iota+\cos \theta \sin \iota},
\end{align}
where $\iota$ is the inclination angle between the magnetic axis and the rotation axis.
%I want to mention here how we compute the derivative, especially the r derivative and its second order, while I realize that this should not be a major concern when we have the analytic form of the field(we can just do the r derivative like we do for theta and phi), The reason why we think determine the r derivative is tricky is we assuming in the trace-back method, we don't acctually know the alpha and beta how to evolve when increasing r. But here since we are not mentioning the trace-back part, it is not neccssary to mention what we did here.

It is straightforward to generalize Equation~\eqref{eq:euler-dipole} to a
  shifted dipole configuration. Define the shifted coordinates as:
\begin{equation}
    x_{s} = x + x_{0},\quad y_{s} = y + y_{0},\quad z_{s} = z + z_{0},
    \label{eq:coord-shift}
\end{equation}
where $(x_0, y_0, z_0)$ is the center of the shifted dipole,$(x, y, z)$ is the Cartesian coordinate with respect to the shifted magnetic
  center, and $(x_{s}, y_{s}, z_{s})$ is the Cartesian coordinate with respect
  to the actual center of the neutron star. Equations~\eqref{eq:euler-dipole}
  are valid in the magnetic body frame $(x', y', z')$, whereas the derivatives in
  Equations~\eqref{eq:jt-hat}--\eqref{eq:jt} now need to be taken with respect
  to the lab frame $(r_{s}, \theta_{s}, \varphi_{s})$.

Note that we have adopted the expression for a non-relativistic magnetic dipole in
  Equations~\eqref{eq:euler-dipole}. \citet{2016ApJ...833..258G} has presented a
  general-relativistic solution of the magnetic dipole in terms of Euler
  potentials. However, once the magnetic dipole is shifted from the center of
  mass of the neutron star, the general relativistic solution no longer applies.
  We have verified that even for millisecond pulsars, the GR solution given
  by~\citet{2016ApJ...833..258G} only differs from the non-relativistic dipole
  solution by around $3\%$. On the other hand, the frame-dragging effect due
  to the rotation of the star, which is the dominant effect of general relativity, is already built into the prescription of
  computing the current density given in
  Equations~\eqref{eq:jt-hat}--\eqref{eq:jt}.

The body frame coordinates $(r, \theta', \varphi')$ are related to the lab
  frame coordinates $(r_{s}, \theta_{s}, \varphi_{s})$ by a nonlinear
  transformation. As a result, taking the derivatives of $\alpha$ and $\beta$
  with respect to $(r_{s}, \theta_{s}, \varphi_{s})$ is quite nontrivial. We overcome
  this difficulty by employing the Automatic Differentiation (AD) technique
  while computing the Euler potentials. AD keeps a graph of mathematical
  operations during the calculation of $\alpha$ and $\beta$, then one can easily
  and accurately take the partial derivatives of these complicated functions
  with respect to the coordinates by walking down the computational graph and
  applying the chain rule. Automatic differentiation allows us to achieve
  machine precision when evaluating the derivatives, without the need to
  introduce a grid. This is especially important in Section~\ref{sec:fitting}
  where we need to evaluate the derivatives a large number of times to carry out
  the Markov Chain Monte Carlo (MCMC) fitting procedure. We use the open source Python automatic differentiation package \verb+jax+\footnote{https://github.com/google/jax} to compute the numerical derivatives of the Euler potentials in the production code.

Given the four-current density across the polar cap, a heating prescription is required to compute the temperature distribution. This heating is intimately related to the dissipation of electromagnetic energy, particle acceleration,
  and the pair production process above the neutron star surface. As shown through
  analytic calculations by of~\citet{2008ApJ...683L..41B} and simulations
  by~\citet{2013MNRAS.429...20T}, particle acceleration at the polar cap depends
  on the ratio $j/c\rho_\mathrm{GJ}$, where $j$ is the magnitude of the three-current density {defined in Equations~\eqref{eq:jr-hat}--\eqref{eq:jph-hat}}, and
  $\rho_\mathrm{GJ}$ is the {Goldreich-Julian charge density, which is equivalent to $J^{\hat{t}}$ defined in Equation~\eqref{eq:jt-hat}.}
  % Goldreich-Julian charge density~\cite{1969ApJ...157..869G}.
  If this ratio is locally greater than 1 (the
  four-current is \emph{spacelike}, $J^2 > 0$) or less than 0 ($j$ and $\rho_\mathrm{GJ}$ have
  opposite signs), these regions can build up a large voltage, accelerating
  particles to very high Lorentz factors, triggering $e^{\pm}$ pair production.
  A fraction of the produced pairs precipate on the polar cap, converting some
  of the energy dissipated in the gap to heat the neutron star surface. On the
  other hand, if $0 < j/c\rho_\mathrm{GJ} < 1$ (the four-current is \emph{timelike}, $J^2 < 0$),
  the plasma can flow along the field line with a mildly relativistic Lorentz
  factor without building up a large voltage~\citep{2013ApJ...762...76C}.

In this paper, we follow the simple prescription used by~\citet{Lockhart2019}, where the surface heating power $P$ is proportional to the magnitude of the three-current density $j$. Equating the heating rate with the radiation power gives a simple relation between current density and surface temperature:
\begin{equation}
T(\theta, \phi) \equiv\left\{\begin{array}{lll}
T_0\left|\boldsymbol{j} / j_0\right|^{1 / 4}, & |\rho|<|\boldsymbol{j}|\text { or } \rho_\mathrm{GJ}j_r < 0\\
0, & |\rho|>|\boldsymbol{j}| 
\end{array}\right.
\label{eq:T-scaling}
\end{equation}
Here the $j_0$ and $T_0$ determine the normalization of current density and temperature on neutron star, and the combination $T_{0}/j_{0}^{1/4}$ can be treated as a fitting parameter. In the subsequent sections, we set this constant to unity, since we are only interested in the shape of pulse profile and this constant only affects the normalization of the X-ray flux. As a result, the temperature values computed in this paper are in arbitrary numerical units.
% the temperature and flux are expressed in arbitrary units, their absolute values are less significant.
% \AC{[ADD A COMMENT: In practice, we actually take this constant to be unity, since in most cases we are only interested in the shape of the pulse profile, while this overall constant only changes the normalization of the overall X-ray flux.]}

% \AC{Pulsar polar cap heating is intimately related to the particle acceleration
%   and pair production process above the neutron star surface. As shown through
%   analytic calculations by of~\citet{2008ApJ...683L..41B} and simulations
%   by~\citet{2013MNRAS.429...20T}, particle acceleration at the polar cap depends
%   on the ratio $J/c\rho_\mathrm{GJ}$, where $J$ is the three-current density and
%   $\rho_\mathrm{GJ}$ is the Goldreich-Julian charge
%   density~\cite{1969ApJ...157..869G}. If this ratio is greater than 1 (the
%   four-current is spacelike) or less than 0 ($J$ and $\rho_\mathrm{GJ}$ have
%   opposite signs), these regions can build up a large voltage, accelerating
%   particles to very high Lorentz factors, triggering $e^{\pm}$ pair production.
%   A fraction of the produced pairs precipate on the polar cap, converting some
%   of the energy dissipated in the gap to heat the neutron star surface. On the
%   other hand, if $0 < J/c\rho_\mathrm{GJ} < 1$ (the four-current is timelike),
%   the plasma can flow along the field line with a mildly relativistic Lorentz
%   factor without building up a large voltage~\citep{2013ApJ...762...76C}.}

The region where $\rho_\mathrm{GJ}j_r < 0$ is called the (volumetric) ``return current'' region which always exists in the force-free solution~\citep[see e.g.][]{2012MNRAS.423.1416P}, independent of the return current sheet along the separatrix between the open and closed field line zones. Previous modeling work such as~\citet{Lockhart2019} mostly focused on the
  spacelike region where $j/c\rho_\mathrm{GJ} > 1$, and ignored the return
  current region where $j$ and $\rho_\mathrm{GJ}$ have opposite signs. In this
  paper, we treat both regions using the same heating model described by
  Equation~\eqref{eq:T-scaling}, in order to asses the importance of modeling
  the return current region. As can be seen in Sections~\ref{sec:results}
  and~\ref{sec:fitting}, the return current region may contribute a
  non-negligible amount to the X-ray light curves, influencing the results of
  the fitting process.

% Only the space-like region, namely $|\rho|<|\boldsymbol{J}|$ could be hotspot region, where the pair production will occur, See more discussion about this in (Timokhin \& Arons 2013; Philippov et al. 2015b).In application, constant $T_0$ could be an adjustable parameters to fit the observed Pulse profile since directly determine this value is challenging and highly atmospheric structure model-dependent. Thus when computing Current density, the $J/J_0$ always be more interesting to demonstrate. Same reason, $T/T_0$ will be illustrated in following sectio
\begin{figure*}
    \centering
    \begin{minipage}{\linewidth}
        \centering
        \includegraphics[width=0.6\linewidth]{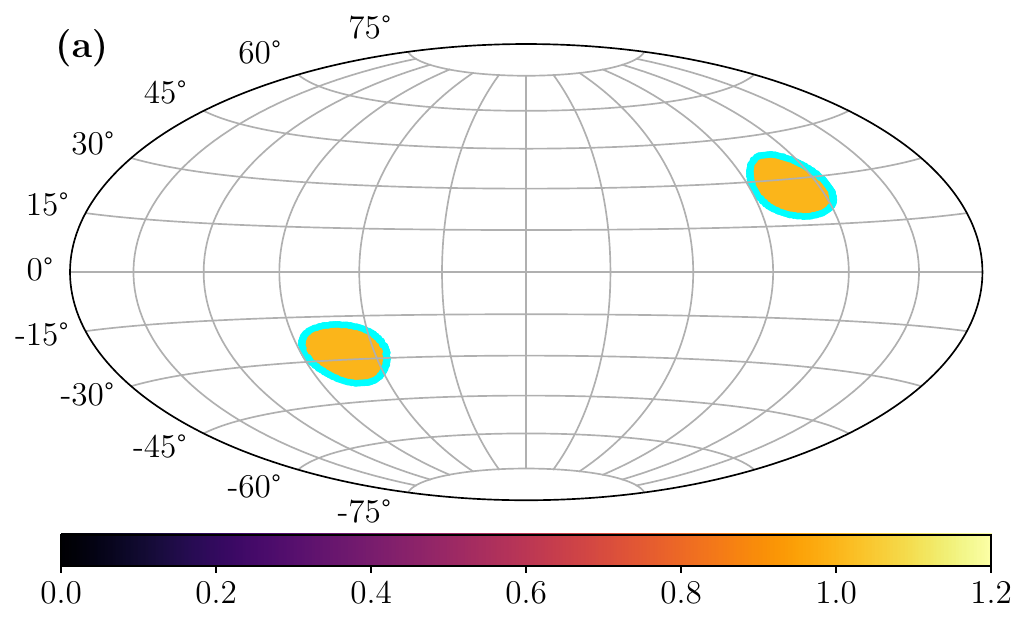}
        \vspace{0.1in}
    \end{minipage}%
    
    \begin{minipage}{\linewidth}
        \centering
        \includegraphics[width=0.6\linewidth]{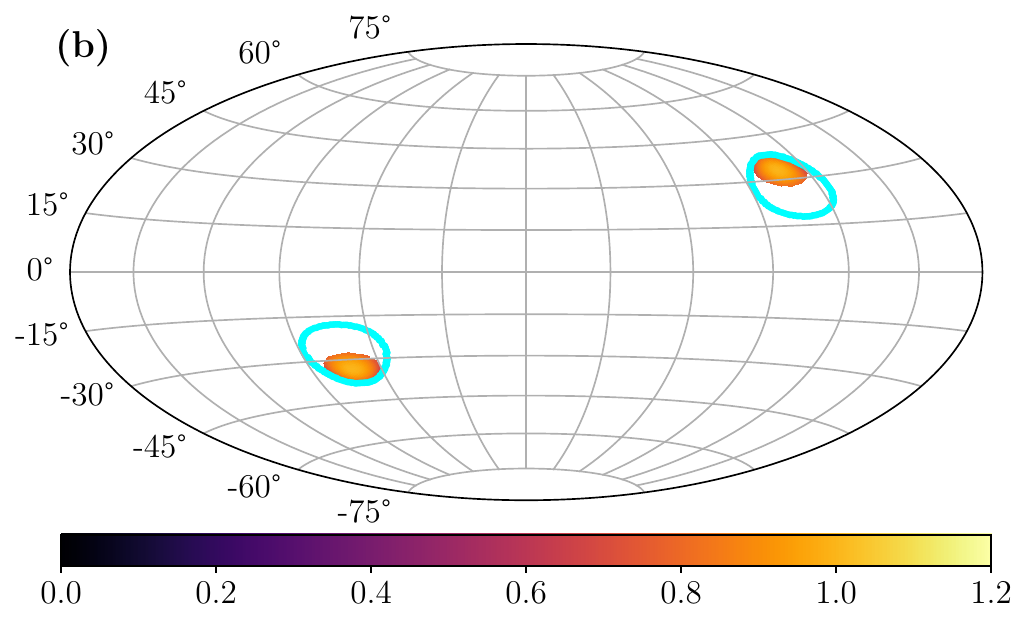}
        \vspace{0.1in}
    \end{minipage}
    \begin{minipage}{\linewidth}
        \centering
        \includegraphics[width=0.6\linewidth]{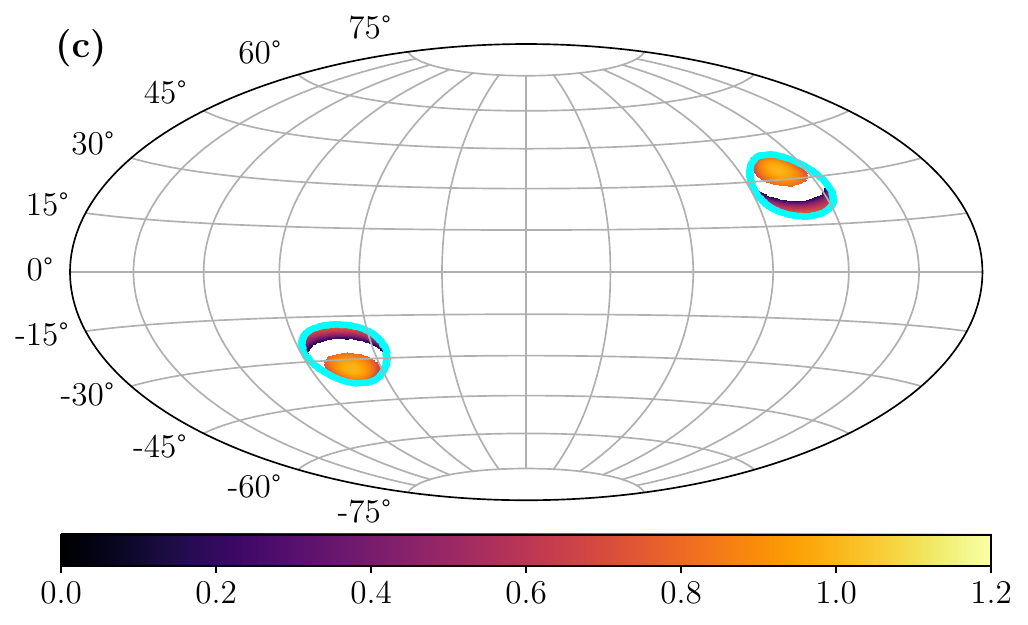}
    \end{minipage}
    \caption{Temperature distribution of different hotspost configurations (magnetic inclination $\iota = 45^\circ$.):(a) Uniform centered dipole circular hotspots with $T = 10^6\,K$.(b) The centered dipole with computed temperature distribution without considering return current. (c) The centered dipole with computed temperature distribution with considering return current region. All the temperature profile normalized by the maximum temperature of this profile. The cyan boundary shared by a,b,c represent the polar cap boundary delineated by $\alpha = \alpha_0$ from Equation~\eqref{eq:alpha0}.}
    \label{fig:shift_canonical}
etnec\end{figure*}
\begin{figure}
\centering
\includegraphics[scale=0.55]{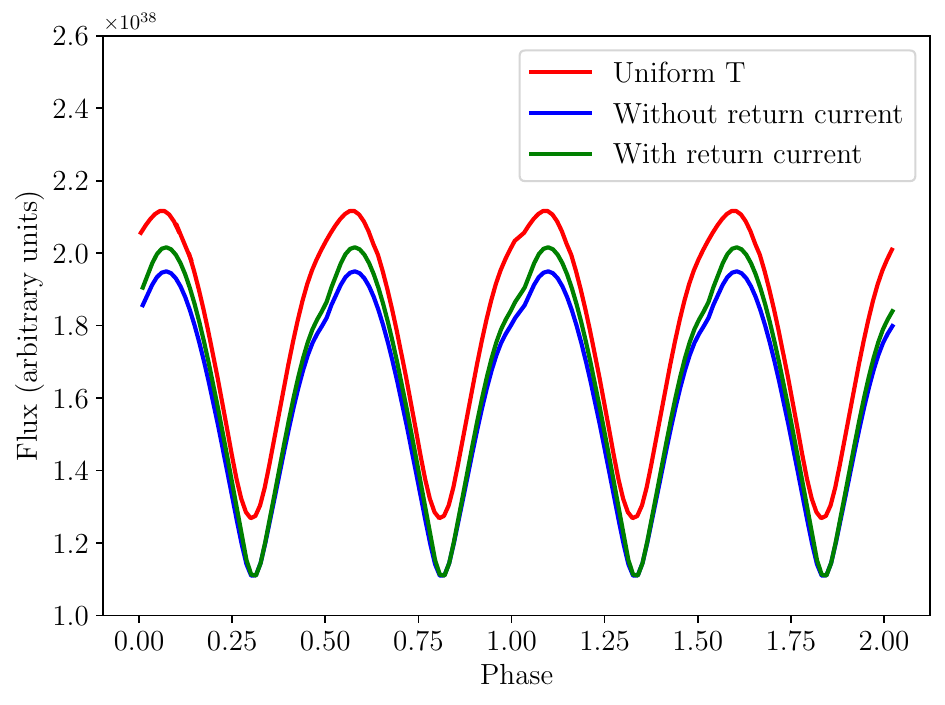}
    \caption{Pulse profile of bolometric flux computed from the centered dipole field in three cases: Uniform hotspot temperature (red), Without return current (blue), With return current (green). All the other parameters to generate this plot are listed in Table \ref{tab1:xpsi_para}. The magnetic inclination angle $\iota=45^{\circ}$, observer inclination angle $I = 90^{\circ}$.}
    \label{fig:shift1dppm}
\end{figure}

% To convert the temperature profile into X-ray emission, the \textit{X-PSI} package, as outlined by Riley et al. (2023), is employed.
To compute the X-ray light curves from the surface temperature profiles, we use the open source \emph{X-PSI} package used by ~\citet{Riley2019} (\url{https://github.com/xpsi-group/xpsi.git}, \citet{xpsi}), which 
follows a methodology developed over many previous works \citep{Riley_phd}. This code was bechmarked in \citet{Bogdanov_2019,Bogdanov_2021}, and used in analysis \citet{Riley2019, Riley_2021}. To describe the shape of the hotspots, the \emph{X-PSI} package normally utilizes a model that represents the hot-spot surface through overlapping spherical caps with uniform temperature, but it also has the capability to accommodate irregular shapes of hot-spots by specifying the temperature distribution over the whole stellar surface ~\citep[see e.g.][]{Das:2024jxc}. It is the later mode that we are using in the calculations presented below. For definitiveness, the physical parameters we used for computing the light curves are summarized in Table~\ref{tab1:xpsi_para}.
% comprehensive full-star surface temperature map.

In the closed field line zone of the pulsar magnetosphere, no current is flowing along the magnetic field lines. As a result, our simple model would predict there is zero heating in these regions which often occupies a large portion of the stellar surface.
For these areas, a uniform temperature of $10^{3}$ in our numerical units is assigned as the background temperature when computing the sample light curve. We keep this surface temperature $T_s$ a fitting parameter for Section~\ref{sec:fitting}. The choice of fitting parameters will be discussed in more detail in Section~\ref{sec:inference-method}.

% {In Section~\ref{sec:fitting}, the parameter $T_0/j^{1/4}$ will not be treated as an independent free parameter but will be incorporated into the fitting of $T_s$. Because the light curve is normalized by its maximum flux value, the ratio of the hotspot temperature to the background temperature predominantly determines the light curve shape. The parameter $T_0/j^{1/4}$ solely affects the absolute value of the hotspot temperature. Therefore, owing to the significant degeneracy between $T_s$ and $T_0/j^{1/4}$ on determining the normalized light curve shape, we fix $T_0/j^{1/4}$ at unity and treat $T_s$ as the fitting constant.}\AC{[TODO: Modify this paragraph to discuss merging $T_s$ and $T_0/j^{1/4}$. ]}

\section{Results}
\label{sec:results}
\subsection{Centered Dipole}
\begin{figure*}
    \begin{minipage}{0.49\linewidth}
    \includegraphics[width=1\linewidth]{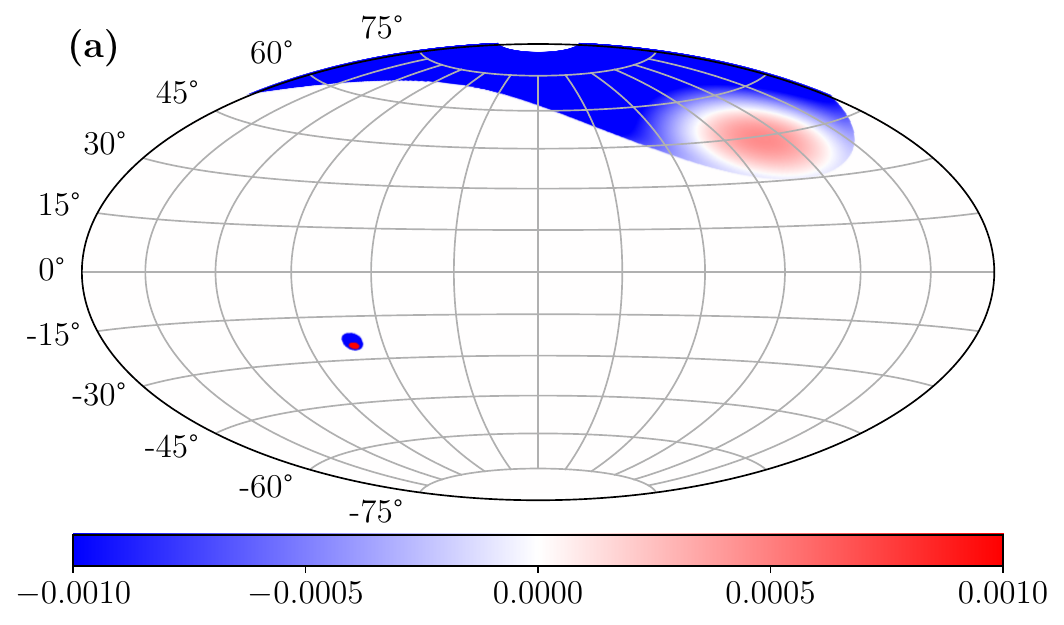}\par
        \label{fig2:xshift_curren}
    \end{minipage}%
    \begin{minipage}{0.49\linewidth}
        \includegraphics[width=1\linewidth]{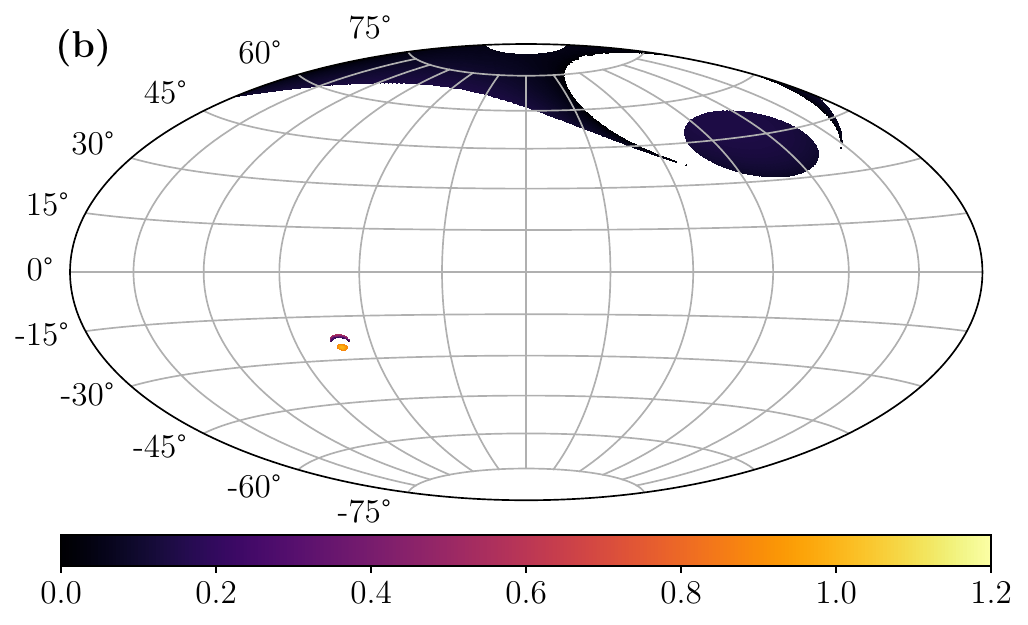}\par
        \label{fig2:yshift_curren}
    \end{minipage}
    \vspace{\floatsep} % Adjust the vertical space between rows
    \begin{minipage}{0.49\linewidth}
        \includegraphics[width=1\linewidth]{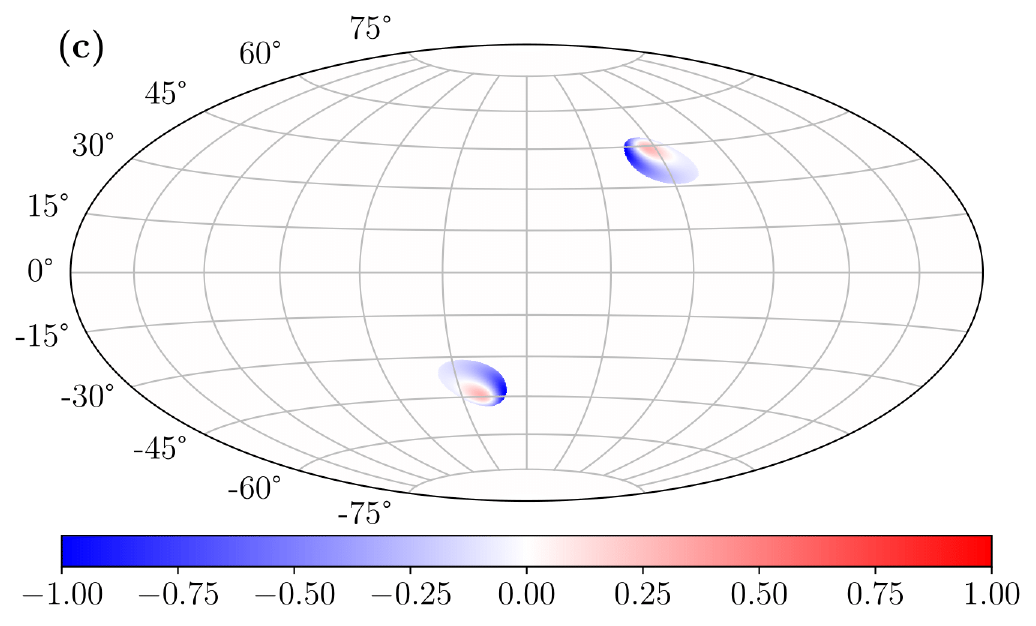}\par
        \label{fig2:zshift_curren}
    \end{minipage}%
    \begin{minipage}{0.49\linewidth}
    \includegraphics[width=1\linewidth]{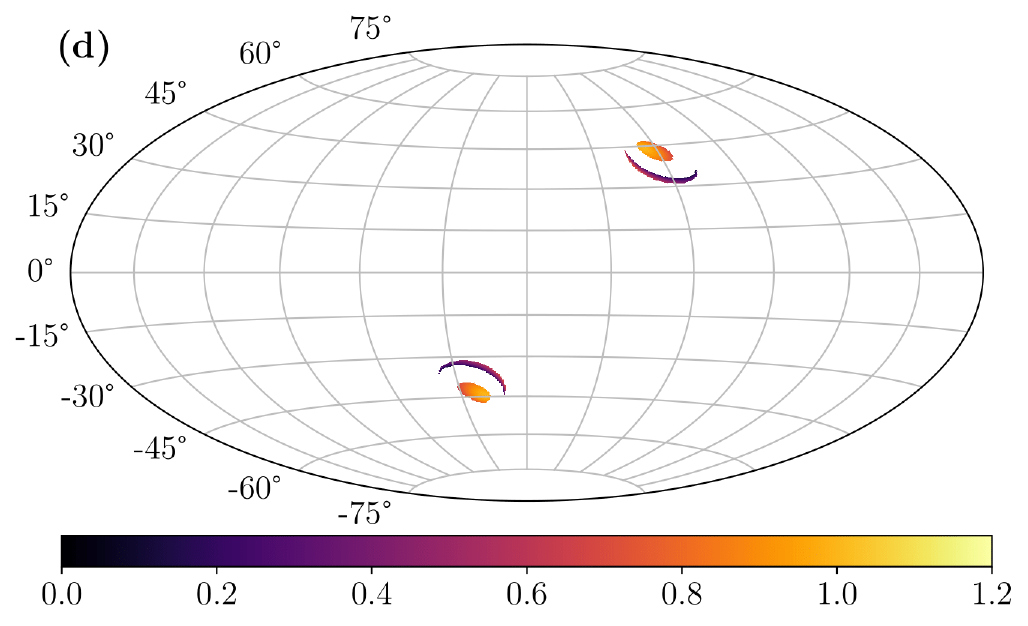}\par
        \label{fig2:xshift_T}
    \end{minipage}%
    \vspace{\floatsep} % Adjust the vertical space between rows
    \begin{minipage}{0.49\linewidth}
        \includegraphics[width=1\linewidth]{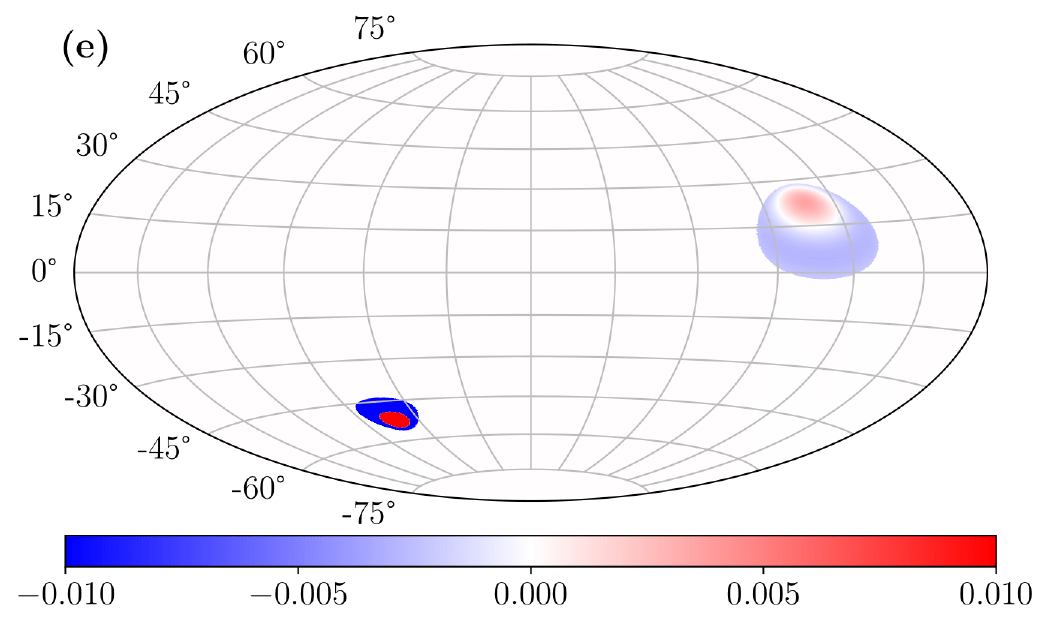}\par
        \label{fig2:yshift_T}
    \end{minipage}
    \begin{minipage}{0.49\linewidth}
        \includegraphics[width=1\linewidth]{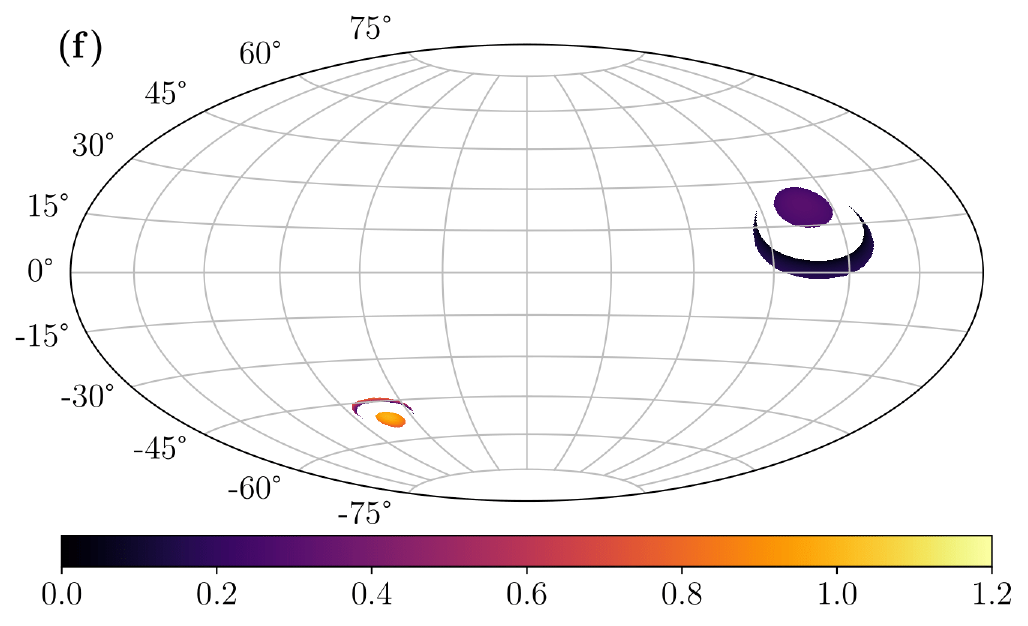}\par
        \label{fig2:zshift_T}
    \end{minipage}%
    
    \caption{4-current density squared $J^{2}$ and temperature $T$ distribution of shifted hotspot projected on the star surface, with magnetic inclination $\iota = 45^{\circ}$ and dipole shift $\delta = R_{\star}/2$ along different directions. Panels (a) and (b): $x$--direction shift; panels (c) and (d): $y$--direction; panels (e) and (f): $z$--direction shift. Each color map is normalized by the maximum value of $J^{2}$ and $T$. 
    % Different figures with different maximum color range for illustrative purposes.
    }
    \label{fig:shiftcombined}
\end{figure*}
\begin{figure}
\centering
\includegraphics[scale=0.5]{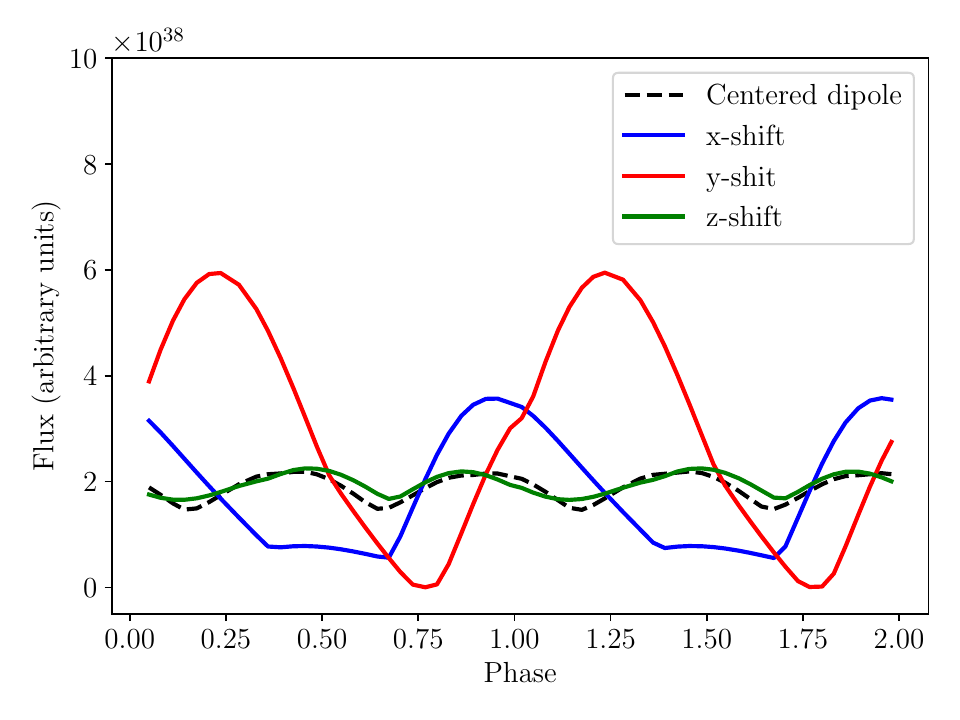}
    \caption{Pulse profile computed from the shifted dipole field in three cases: $x$--direction shift (blue), $y$--direction shift (red) and $z$--direction shift (green), comparing with centered dipole field computation without including any shift (black dashed). All the other parameters to generate this plot are listed in Table~\ref{tab1:xpsi_para} with magnetic inclination $\iota = 45^{\circ}$, observer inclination $I = 90^{\circ}$}
    \label{fig:xyzshift1dppm}
\end{figure}
In order to validate our method, we first compute the polar cap temperature distribution for a centered magnetic dipole. Figure~\ref{fig:shift_canonical} compares the temperature distribution for a centered dipole at inclination angle $\iota = 45^\circ$, using three different methods for computing the hotspot temperature distribution: uniform temperature of $T = 10^6\,K$, only spacelike region ($j/c\rho_\mathrm{GJ} > 1)$, and spacelike region plus volumetric return current ($j/c\rho_\mathrm{GJ} < 0$). The shapes of the activated polar caps agree with previous studies such as \citet{2010ApJ...715.1282B,Gralla2017}. We find that the volumetric return current region always produces a small crescent-shaped region with much lower temperature than the main polar cap. 
% In contrast to the analyses presented in \citet{Gralla2017} and \citet{Lockhart2019}, the light curves in this study have not been normalized, as our aim is to display their absolute magnitudes. \AC{As a result, the uniform temperature case is has overall higher flux due to a larger emitting area.} Nonetheless, the overall morphology of the light curves is consistent with the findings reported in those studies. 
% reproduce the canonical dipole field introduced Temperature profile of neutron star.
% Variations in the inclination angle of the dipole lead to differing temperature mappings, as elucidated in the preceding section.

% In Figure \ref{fig:shift_canonical}, we illustrate three different canonical dipole configurations. The first configuration shows uniform temperature hotspots with an inclination angle of $\pi/3$. The second configuration is the hotspot arrangement computed using the method described above, also with an inclination angle of $\pi/3$ in the canonical dipole regime. The third configuration considers the return current region as an additional X-ray emission region, which was neglected by Gralla (2017) and Lockhart (2019). In the canonical dipole case, ignoring this return current region is reasonable since its emission contribution is substantial.

We also computed the pulse profiles of each case assuming an observer at observer inclination $I = 90^\circ$. The results are shown in Figure~\ref{fig:shift1dppm}. Similar to \citet{Lockhart2019}, we do not observe significant deviation between all three pulse profiles, The elevated amplitude observed in the light curve derived from the uniform temperature computation is primarily attributable to our specific choice of temperature. Determining an appropriate uniform temperature that accurately represents the underlying physics is challenging; therefore, we have adopted a value of $10^6\,K$ solely for demonstrative and comparative purposes. 
% The uniform temperature light curve leads to an overall higher flux simply due to its larger emission area. 
% In our realistic hotspot computations, the temperature profiles are expressed in arbitrary units. 
Although the uniform temperature model produces a light curve very similar to our model with more realistic hotspot configuration, it is but a coincidence in our choice of parameters.
% comparing the absolute flux values between the realistic hotspot (with or without return current) and the uniform temperature model is not meaningful. This comparison is presented solely to demonstrate the overall similarity in the light curve shapes across different models under centered dipole regime and to highlight subtle differences between each setup.

In particular, Figure~\ref{fig:shift1dppm} shows that the contribution of the volumetric return current region is quite subdominant, justifying the choice of ignoring this component in previous studies. However, in our subsequent investigations of a shifted dipole magnetic configuration, there is no reason \emph{a priori} to expect the return current continues to play a subdominant role. Therefore, we will keep this component in all subsequent calculations to assess its importance.

\subsection{Shifted Dipole}
\label{sec:shifted-dipole}

We now apply the methods described in Section~\ref{sec:method} to compute the temperature maps and X-ray light curves of shifted dipole configurations. In this section, our goal is to demonstrate the feasibility of our method, only exploring the impact of dipole displacement on the current and temperature distribution for specific parameters. In particular, we examine a magnetic dipole shifted in $x$, $y$, and $z$ direction respectively, each by a distance of $\delta = R_*/2$.
This shift is intentionally chosen to be large, in order to show its dramatic effects on both the current density and temperature profiles, as well as on the X-ray pulse profile.

Figure \ref{fig:shiftcombined} illustrates the effect of a shift $\delta = R_{\star}/2$ along the $x$, $y$, and $z$ directions on the polar cap 4-current density structure and temperature distribution. For example, in the panel (a), we present the $x$--direction shift field 4-current density distribution across the polar cap on the stellar surface. Compared to the centered dipole configuration, the north polar cap is greatly extended to a larger area, with a single hotspot at its center. In contrast, the south polar cap is compressed into a much smaller region. It should be noted that the small blank region at the center of the north pole is an artifact of the projection used in the plot.
% polar plotting.
In reality, the hot spot does not exhibit any hollow feature. The panel (b) shows that the south polar cap , despite being much smaller, has a significantly higher temperature than the north. {This follows from the fact that, in our model,
the local surface temperature is set by the magnitude of the
current density $j$. Each
open magnetic flux surface carries a conserved current
$\Lambda(\alpha,\beta)$, fixed in the far zone where the field is
symmetric across the current sheet.  Because both caps intercept the
same total current, squeezing one cap into a smaller area inevitably
raises $j$ there, and thus (after the red-shift factor
$\sqrt{1-2M/R_\star}$ is applied) delivers more heating per unit
area.} Similar asymmetry between the temperatures of the two polar caps can be seen in the $z$--shifted dipole case as well (see panels (e) and (f) of Figure~\ref{fig:shiftcombined}).

The four-current density and temperature distribution for the $y$--shift scenario $\delta_y = R_{\star}/2$ in a dipole field are shown in the panels (c) and (d) of Figure~\ref{fig:shiftcombined}. The two polar caps remain symmetric, and generally similar to the unshifted case shown in Figure~\ref{fig:shift_canonical}. The main difference is that the hot spots become closer together compared to the centered dipole scenario. This displacement changes the distance between the two peaks in the X-ray light curve in one rotation period, as can be seen in Figure~\ref{fig:xyzshift1dppm}.
% and the resultant changes in the pulse profile emission peaks imply that observational data could strongly constrain the magnitude of the $y$-shift. 
Such a distinct outcome for a $y$--direction shift arises naturally because the star's dipole moment lies in the $x-z$ plane, so shifting perpendicular to that plane yields behavior not seen in within-plane shifts.

% The panel (c) displays the current density distribution when the shift is along the $y$-direction, producing a circular hotspot at the edge of the polar cap. The panel (d) illustrates the temperature field derived from this current density. From the global stellar map, it is evident that both hotspots exhibit similar temperatures. In this $y$-shift scenario, the hotspots move closer together compared to the $x$-shift case. 

The panels (e) and (f) of Figure~\ref{fig:shiftcombined} show how a shift $\delta_z = R_{\star}/2$ along the $z$--direction affects the hotspot current density distribution and temperature map. In panel (e), the two polar caps display noticeably different areas, with the north polar cap undergoing an enlargement similar to what was observed in the $x$--shift scenario, and southern polar cap also exhibited a larger current density similar with what we see in $x$--direction shift. The panel (f) reveals that, in this configuration, the south hotspot appears significantly brighter, whereas the north hotspot---despite having a larger area--- is comparatively fainter, similiar to the features seen in the $x$--shift case.

The X-ray pulse profiles for shifts in the $x$--, $y$--, and $z$--directions are computed and presented in Figure~\ref{fig:xyzshift1dppm} for a neutron star with magnetic inclination angle $\iota = 45^{\circ}$, as seen by an observer at angle $I = 90^\circ$ with respect to the rotation axis. Additional parameters, such as the mass and radius of the neutron star, are required to determine its external spacetime, thereby allowing us to generate the pulse profiles using the \emph{X-PSI} package. We used the parameters of the pulsar J0030+0451 for these light curves, and the values are summarized in Table~\ref{tab1:xpsi_para}.
% Table~\ref{tab1:xpsi_para} lists these parameters, here in this computation we use J0030 paramaters as standard input to compute light curve (including its mass, radius, distance to the source, and rotation frequency $\nu$). Once these factors are specified, the resulting pulse profile becomes uniquely determined.

Without any normalization added, our model indicates that for an $x$--direction shift, X-ray emission from the south polar cap dominates, yielding a significantly larger peak-to-valley ratio than the centered-dipole light curve (shown as the dashed curve). Although the south polar cap is smaller in area, it is brighter due to the locally enhanced magnetic field. Consequently, the maximum peak in the $x$--shift configuration is nearly twice that of the centered-dipole setup. In the $y$--direction shift scenario, the north and south polar caps move closer, merging into a single, highly boosted emission peak; Figure~\ref{fig:xyzshift1dppm} shows that this configuration produces both the highest peak and the largest peak-to-valley ratio, implying a higher total flux than in the other shift cases. Finally, the $z$--direction shift presents an interesting contrast. Although the north polar cap appears fainter than the south cap in Figure~\ref{fig:shiftcombined}, the observer inclination $ I = 90^\circ$ (equatorial viewing) brings the north cap closer to the line of sight. As a result, the pulse profile exhibits the smallest peak-to-valley ratio, and the south cap’s higher temperature does not substantially exceed the emission level seen in the centered-dipole configuration.

\begin{table*}[ht]
\centering
\setlength{\tabcolsep}{9mm}{\begin{tabular}{l c c} 
\hline\hline
\text {Model Parameters} & {PSR~J0030+0451} & {PSR~J0437--4715} \\
\hline
Mass $M_{\star}$ & 1.385 $M_{\odot}$ & 1.418 $M_{\odot}$\\ Radius $R_{\star}$  &11.70 km &11.36 km \\
Earth distance $D$ & 0.2 kpc & 156.98 pc\\
$\nu$& 205.3 Hz & 174 Hz\\ 
$B$ & $5\times10^{8}$ G & $5\times10^{8}$ G
\\
\hline\hline
\end{tabular}}
\caption{This is a summary for all the pulse-profile modeling parameters setting for (a) PSR J0030+0451 and (b) PSR J0437--4715. The J0030 parameters are also used in Section~\ref{sec:results} to compute the hot spot temperature distribution of centered and shifted dipoles. Mass $M_{\star}$ Radius $R_{\star}$ refers the neutron star mass and radius, earth distance $D$ defined as the distance between this source and earth, $\nu$ is the Coordinate frequency of the mode of radiative asymmetry in the photosphere that is assumed to generate the pulsed signal [Hz]. $B$ in G [Gauss], is the surface magnetic field that take into the computation of magnetic moment. The frequency and earth distance of J0030 are derived from radio observation \citet{Arzoumanian_2018}, and the frequency and earth distance of J0437 was from \citet{Reardon_2024}.
}
\label{tab1:xpsi_para}
\end{table*}

\begin{table}[ht]
\centering
\setlength{\tabcolsep}{7mm}{\begin{tabular}{l c} 
\hline\hline
\text {Model Parameters} & {Prior} \\
\hline
observer inclination $I$& $\mathcal{U}(0, \frac{\pi}{2})$\\
magnetic inclination $\iota$ & $\mathcal{U}(0, \frac{\pi}{2})$ \\
$x_0$ & $\mathcal{U}(-R_{\star}, R_{\star})$\\ 
$y_0$ & $\mathcal{U}(-R_{\star}, R_{\star})$\\
$z_0$ & $\mathcal{U}(-R_{\star}, R_{\star})$\\
$\log_{10}T_{s}$ &  $\mathcal{U}(0,10)$\\

\hline\hline
\end{tabular}}
\caption{Summary of Inference Model Parameters and Prior Settings: The observer inclination is denoted by $I$ and the magnetic inclination by $\iota$. The off-center dipole displacements along the $x$--, $y$--, and $z$--axes are represented by $x_0$, $y_0$, and $z_0$, respectively. $T_s$ denotes the temperature of the neutron star's non-hotspot region (in arbitrary units). Here, $\mathcal{U}$ denotes a uniform distribution.}
\label{tab2:inferened_para}
\end{table}

\begin{figure*}[ht]
\centering
\includegraphics[scale=0.55]{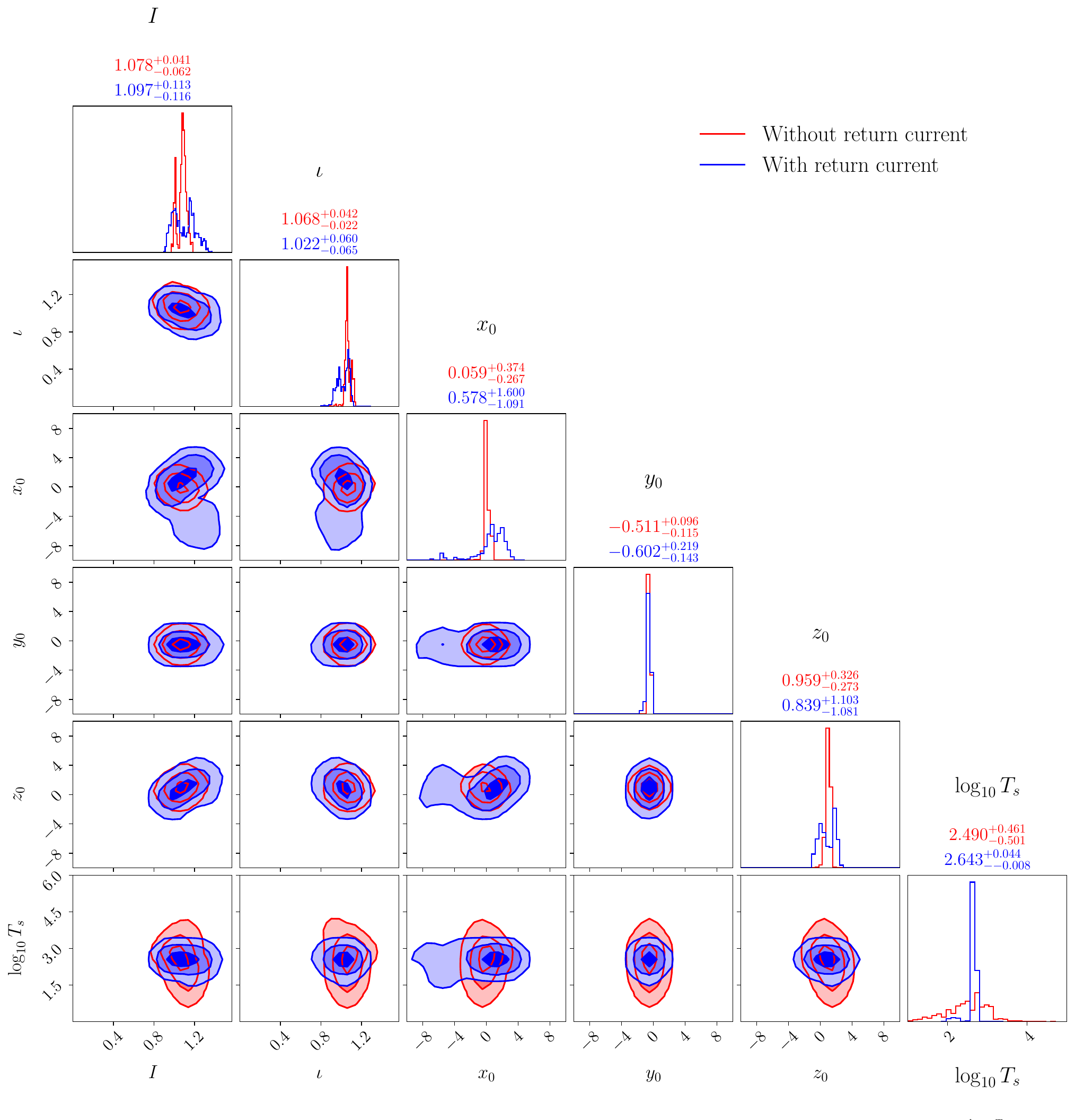}
    \caption{The posterior of the observer inclination $I$, magnetic inclination angle $\iota$ (both in radians), $x$--$y$--$z$ direction shifts and $\log_{10}T_{s}$ after applying constraints to the observed pulse profile of J0030. Blue is the posterior with Case 1: without return current. Red posterior is result in Case 2: with return current. The contour levels in the corner plot, going from deep to light colours, correspond to the 68\%, 84\% and 98.9\% levels. The title of this plot indicates the median of the distribution as well as the range of the 68\% credible interval. Here, $x$--$y$--$z$ shifts are measured in $\mathrm{km}$, while $T_s$ is in numerical units.}
    \label{fig3:posterior}
    \vspace*{-120ex}  % Tune this to the image height.
    \begin{flushright}
      \begin{tabular}{ccc}
        \footnotesize
        \textbf{Maximum Likelihood} & \textbf{Case 1} & \textbf{Case 2} \\ \hline\hline
        observer inclination $I$ & {0.78} & {0.68} \\ \hline
        magnetic inclination $\iota$ & {1.52} & {0.98} \\ \hline
        $x_0\,(\mathrm{km})$ & {-0.35} & {0.66}  \\ \hline
        $y_0\,(\mathrm{km})$ & {-0.22} &  {-0.29}\\ \hline
        $z_0\,(\mathrm{km})$ & {-0.25} &{-0.16} \\ \hline
        $\log_{10}T_{s}$ & {2.40} & {2.62} \\ \hline
      \end{tabular}
    \end{flushright}
    \vspace*{105ex}  % The spacing above but without the minus.
\end{figure*}
\begin{figure*}
    \centering
    \begin{minipage}{\linewidth}
        \centering
        \includegraphics[width=0.7\linewidth]{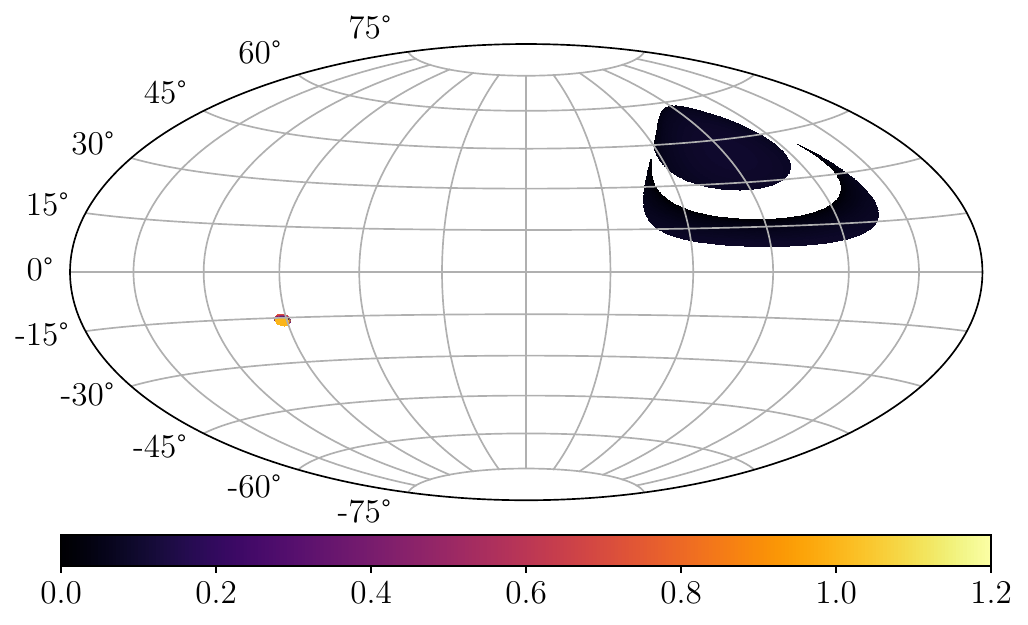}
        \label{3d_hotspot}
    \end{minipage}%
    \begin{minipage}{\linewidth}
        \centering
        \includegraphics[width=0.7\linewidth]{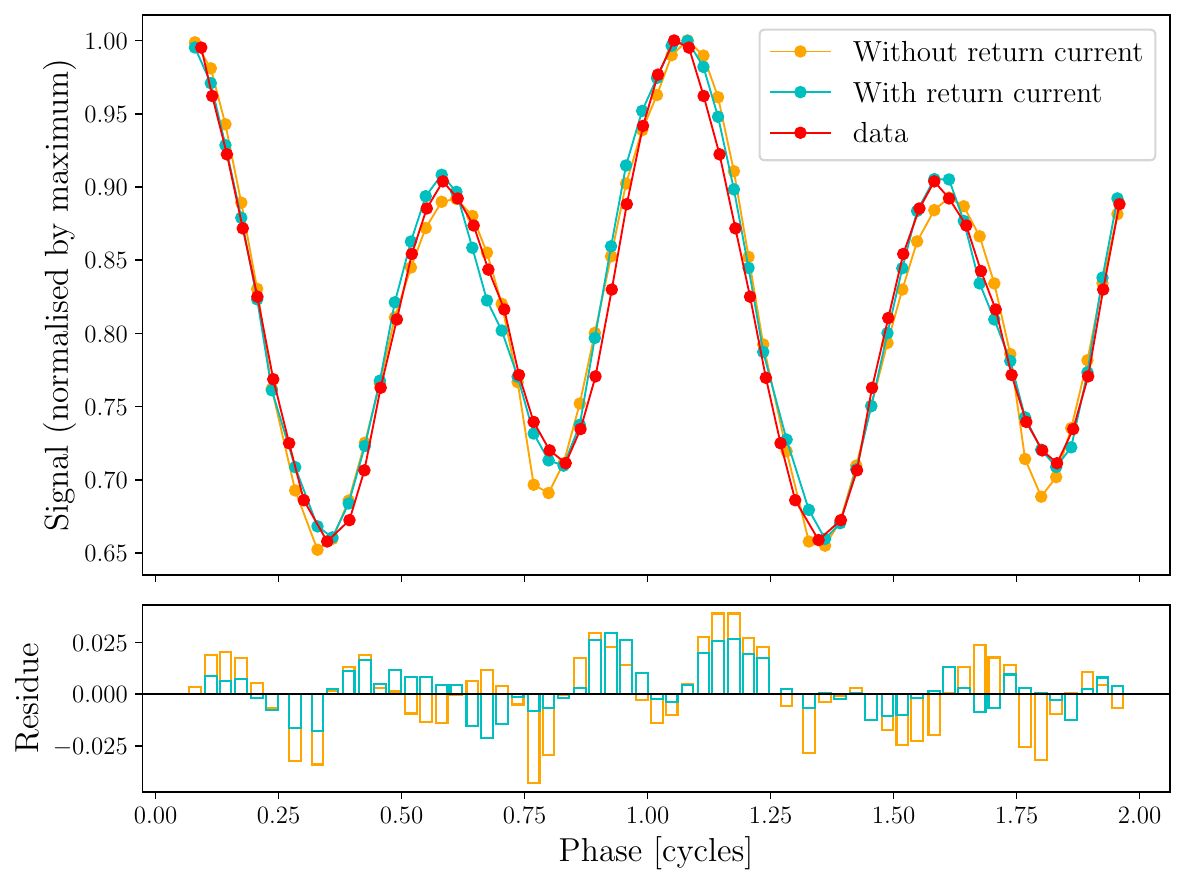}
        \label{pulse_profile}
    \end{minipage}
    \caption{(a) The maximum-likelihood hotspot configuration in Case 2: with return current, the color scale indicating the temperature. (b) The maximum-likelihood parameters produced pulse profile  of J0030 results comparing with the data}
    \label{fig:j0030_profile}
\end{figure*}

\section{Real case study: PSR J0030+0451 and PSR J0437--4715 hotspot map construction}
\label{sec:fitting}

In the sections above, we computed physics-motivated hotspots on the neutron star with an off-center dipole magnetic configuration and derived the temperature distribution of these hotspots. We want to test how well the off-center dipole model can reproduce some of the X-ray observations, and use the results to assess whether higher-order multipole moments are necessary to fully account for the pulsar's X-ray emission.

% The off-center dipole can be viewed as a preliminary approximation of the neutron star's magnetic field, with complexity of introducing multipole moments omitted. The ability to explain observational data with this approximation is already of significant interest. How well this off-center dipole model can explain the data indicates the extent to which higher-order multipole moments are necessary to fully account for the pulsar's X-ray emission.

In this case study, we focus on two sources: PSR J0030+0451 (J0030) and PSR J0437--4715 (J0437). PSR J0030+0451 was the first millisecond pulsar that NICER source for which detailed pulse profile modeling was performed.
% reported to be a $\sim$1.4 $M_{\odot}$ star. 
The original report on the hotspot configuration of this star, by \citet{Riley2019} and \citet{Miller2019}, described one circular hotspot combined with an elongated hotspot, both in the same hemisphere, suggesting the presence of higher-order multipole moments. However, with new background constraint from XMM-Newton data, \citet{Vinciguerra2024} carried out a follow-up study, and concluded that higher-order multipoles may not be necessary, but a dipole with nontrivial polar cap temperature distribution might be sufficient to explain the data. Whether higher-order multipoles are preferred in the inference method described here for this source is thus of great interest.

The other source, PSR J0437--4715, is the brightest source currently available. Recent analysis of its hotspot configuration by \citet{Choudhury_2024} reports a ring-shaped hotspot plus a circular hotspot, which clearly indicates a multipole feature. Investigating whether this shifted dipole field can adequately recover the data for PSR J0437--4715 is also a very intriguing problem.

\subsection{Inference methodology}
\label{sec:inference-method}

\begin{figure*}
\centering
\includegraphics[scale=0.55]{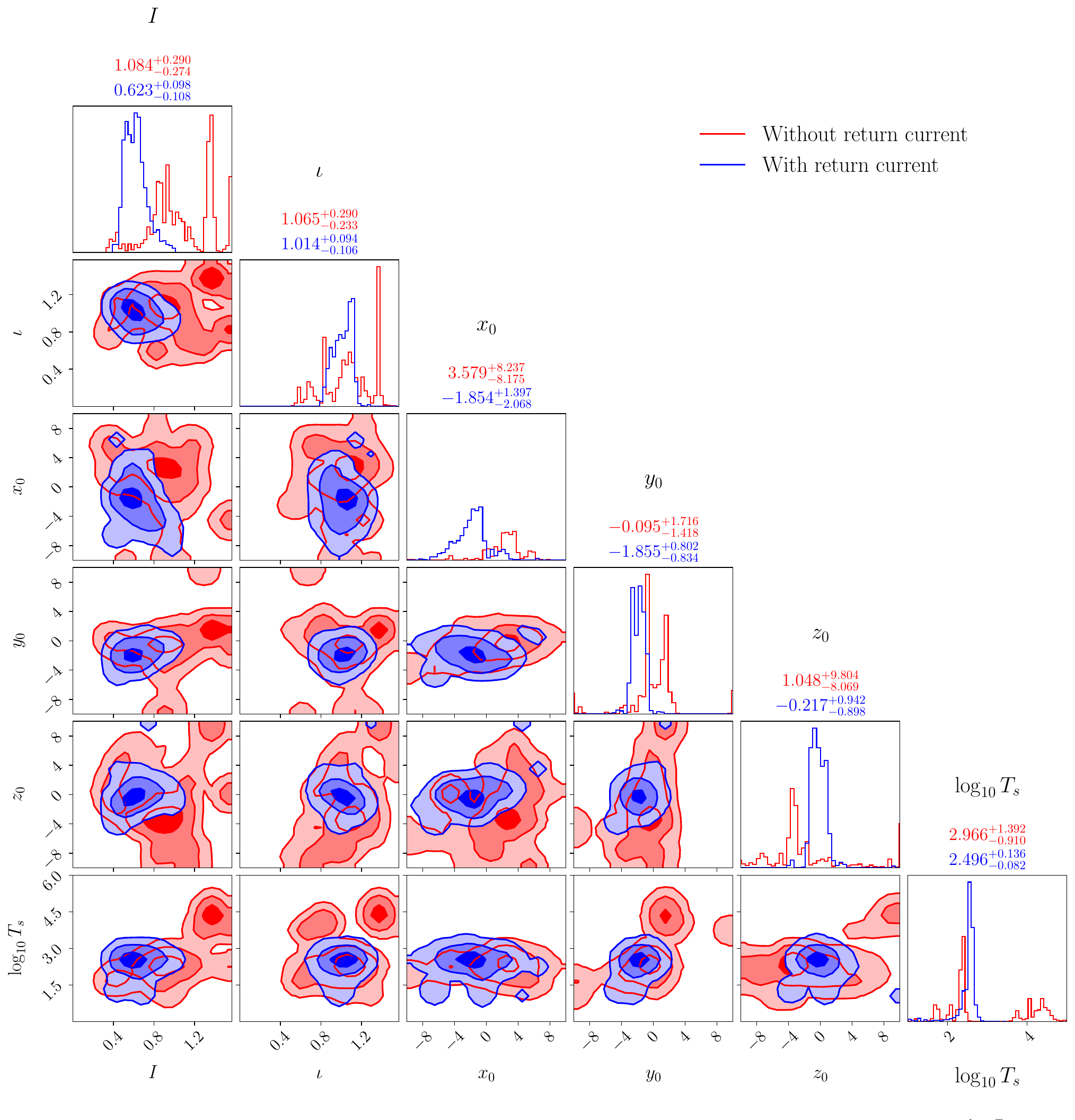}
    \caption{The posterior of the observer inclination $I$, magnetic inclination angle $\iota$ (both in radians), $x$--$y$--$z$ direction shift and $T_{s}$ after applying constraints to the observed pulse profile of J0437. Blue is the posterior with Case 1: without return current. Red posterior is result in Case 2: with return current. The contour levels in the corner plot, going from deep to light colours, correspond to the 68\%, 84\% and 98.9\% levels. The dashed line in the 1D corner plots represents the 68\% credible interval, and the title of this plot indicates the median of the distribution as well as the range of the 68\% credible interval. Here, $x$--$y$--$z$ shifts are measured in $\mathrm{km}$, while the temperature $T_{s}$ is in numerical units}
    \label{fig3:posterior_j0437}
    \vspace*{-120ex}  % Tune this to the image height.
    \begin{flushright}
      \begin{tabular}{ccc}
        \footnotesize
        \textbf{Maximum Likelihood} & \textbf{Case 1} & \textbf{Case 2} \\ \hline\hline
        observer inclination $I$ & {0.40} & {0.59} \\ \hline
        magnetic inclination $\iota$ & {1.22} & {1.09} \\ \hline
        $x_0\,(\mathrm{km})$ & {1.90} & {-1.85}  \\ \hline
        $y_0\,(\mathrm{km})$ & {-0.67} &  {-2.07}\\ \hline
        $z_0\,(\mathrm{km})$ & {-0.27} &{-0.29} \\ \hline
        $\log_{10}T_{s}$ & {2.33} & {2.53} \\ \hline
      \end{tabular}
    \end{flushright}
    \vspace*{103ex}  % The spacing above but without the minus.
\end{figure*}

% We describe the inference procedure as follows: The NICER-informed 1-d marginal pulse profile serves as the observable to constrain the parameters of the shifted dipole field.
We carry out Bayesian inference using the Markov Chain Monte Carlo (MCMC) method to find the best-fitting parameters for a shifted dipole model that can best reproduce the observed X-ray pulse profiles of the two target sources.
To simplify this problem, we fix the mass and radius of these two sources to the maximum likelihood values reported in recent literature.
% selected from NICER posterior samples.
For J0030, these values are $M = 1.385 M_{\odot}$ and $R = 11.70\,\mathrm{km}$, as suggested by the maximum likelihood value from ST-PDT case by \citet{Vinciguerra2024}. For J0437, the values are $M = 1.418 M_{\odot}$ and $R = 11.36\,\mathrm{km}$, which corresponded to the result reported by \citet{Choudhury_2024}. We summarize these fixed model input value in Table \ref{tab1:xpsi_para} for J0030 and J0437 separately. This approach effectively fixes the space-time properties of the pulsar, allowing us to focus solely on the influence of the pulse profile introduced by the hotspot configuration, greatly reducing the parameter space and the amount of computational power required. {The MCMC chains presented below are exploratory just sufficient to demonstrate that the
physics-motivated\,/\,shifted-dipole model presented here can be sampled with a standard Bayesian pipeline.  We have not yet established statistical convergence across alternative sampling schemes, even though the chains display qualitative convergence indicators, and Figures~\ref{fig3:posterior_j0437} still show minimal visible sampling artifacts despite cosmetic changes (wider axes and KDE contours) adopted
to minimize distraction.  A full-scale analysis with longer chains,
multiple sampler configurations, and the mass–radius parameters set free is deferred to future work.}

%The observer inclination is also fixed: for J0030, $\cos(I) = 0.124$, and for J0437, $\cos(I) = -0.7373$. 
The free parameters need to be explored include the the observer inclination $I$, magnetic inclination angle $\iota$, and the $x$--, $y$--, $z$-- direction shifts in the magnetic field: $x_0$, $y_0$, and $z_0$. Since our model predicts only the temperature of the hotspots, we must also account for the temperature of the star's surface outside the hotspot regions. We assume this surface temperature, $T_{s}$, to be uniform and finite across the whole star surface. This parameter is also free and needs to be explored by our inference. 

In principle, the normalization constant of surface temperature, $T_0/j_0^{1/4}$, should be included as a fitting parameter as well. However, since we only use the normalized 1D pulse profile as our MCMC input, discarding the X-ray spectra, the shape of the light curve only depends on the ratio $T_0/T_s$. As we have chosen $T_0/j_0^{1/4}$ to be unity for simplicity, having $T_s$ as a free parameter is enough degrees of freedom.
% {As discussed above, $T_s$ effectively absorbs the effect of tuning the value of $T_0/j^{1/4}$, as only the ratio of these two quantities influences the light curve shape. Consequently, $T_s$ is expressed in arbitrary units and does not reflect the actual neutron star surface temperature ---- its value varies with the choice of the constant $T_0/j^{1/4}$. This degeneracy can only be resolved by simultaneously considering both energy and timing information from the X-ray pulse profile. Since we are fitting a normalized 1-d light curve, introducing an additional free parameter for $T_0/j^{1/4}$ would lead to overparameterization.} \AC{[TODO: Add discussion about the meaning of this surface temperature, and potentially its unit.]}

The prior settings for all free parameters are summarized in Table \ref{tab2:inferened_para}. All six free parameters are set with a uniform prior (denoted as $\mathcal{U}$) to reflect the fact that we do not have a clear preference for any specific values. The $x$--, $y$--, $z$-- direction shifts have a potential constraint: the total shift of the dipole field should be smaller than the star's radius. The likelihood of the simulated light curve compared to the real data is expressed as a mean square difference error, reflecting how closely the simulated pulse profile matches the observed data. By combining the prior and likelihood, we can extract the posterior distribution and explore the allowed parameter space. 

All the computations of X-ray light curve are based on the \emph{X-PSI} package \citep{xpsi}, incorporating a newly developed global temperature map module. This module theoretically handles arbitrary shapes of hotspots and temperature distributions, taking into account effects such as general relativistic light bending and rotational Doppler shift. Here we used blackbody as the emission model without including realistic atmosphere model to reduce the computation time. A similar module has been implemented in \citet{Das:2024jxc}. The MCMC fitting is performed using the \texttt{emcee} software \citep{emcee}.

\begin{figure*}
    \centering
    \begin{minipage}{\linewidth}
        \centering
        \includegraphics[width=0.7\linewidth]{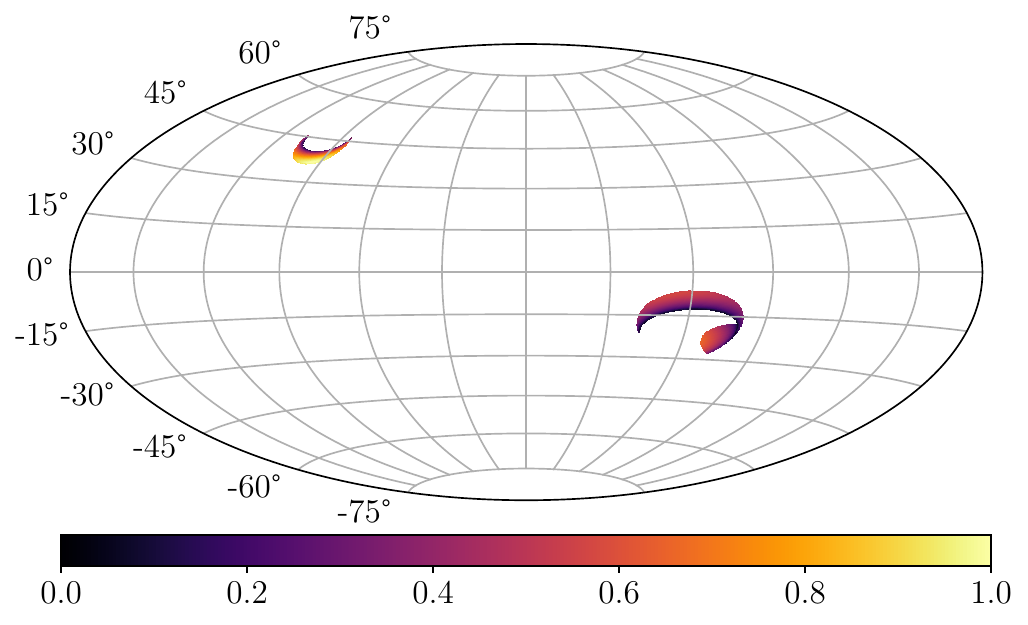}
        \label{3d_hotspot_J0437}
    \end{minipage}%
    \begin{minipage}{\linewidth}
        \centering
        \includegraphics[width=0.7\linewidth]{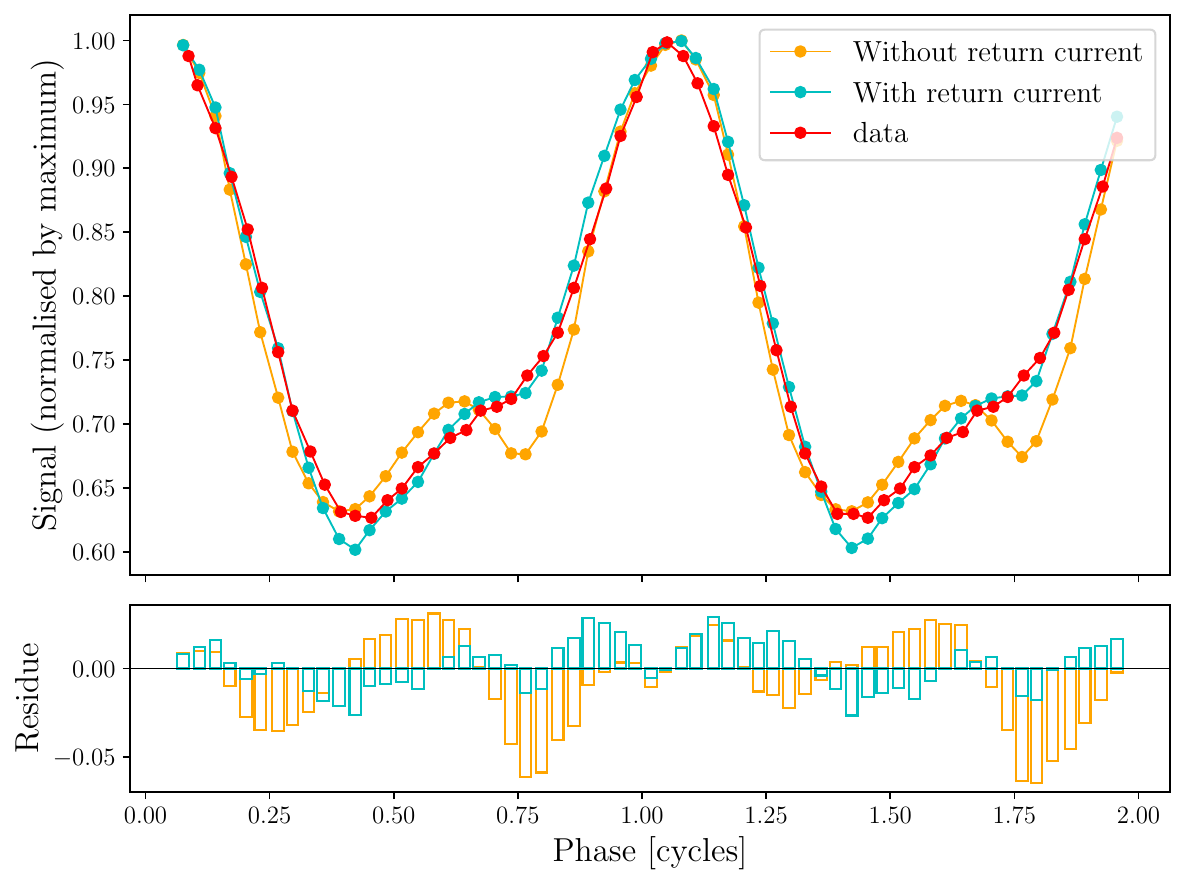}
        \label{pulse_profile_J0437}
    \end{minipage}
    \caption{(a) Polar cap temperature distribution for the maximum-likelihood hotspot configuration in Case 2: with return current for PSR J0437--4715. (b) The maximum-likelihood parameters produced pulse profile  of J0437 results comparing with the bolometric flux data in NICER band. {A companion animated figure showing the evolution of this hotspot configuration as the off‑center distance increases toward the best‑fit state is available online; an archived copy is hosted on Zenodo: \dataset[doi:10.5281/zenodo.16548609]{https://doi.org/10.5281/zenodo.16548609}.}
}
    \label{fig:shiftcombined_J0437}
\end{figure*}

In these analyses, we set up two cases to explore the effects of the volumetric return current, which is typically neglected in the standard dipole configuration of \cite{Lockhart2019}. As discussed above, ignoring this effect leads to a slightly altered hotspot configuration and omits one potential hotspot component. Since its impact has not been proved to be negligible in an off-center dipole scenario, we quantify its influence by performing Bayesian inference under each cases and comparing the resulting maximum-likelihood parameters.

\subsection{Inference results: PSR J0030+0451}

Figure~\ref{fig3:posterior} shows the J0030 full posterior distributions of the model parameters. Whether we perform the fitting without the return current (Case~1) or with return current (Case~2), most parameters converge to pronounced peaks in the posterior space. The observer inclination angle exhibits a consistent peak around $1.0\, \mathrm{rad}$ in both cases, suggesting that this geometry is robust to the presence or absence of return current effects. { The posterior distribution of the magnetic inclination angle, $\iota$, converges to a single dominant peak near $1.0$, indicating satisfactory convergence even without repeated runs under alternative sampling configurations.}

{For the $x$-, $y$-, and $z$-shift parameters, the posterior distributions indicate that, for J0030+0451, the stellar dipole is displaced by only a small fraction of the stellar radius: all shifts are well constrained within 1 km (i.e., $<10\%$ of $R_\star$).
} In terms of the surface temperature outside the hotspot, $T_{s}$, both cases yield similar peak values, but the uncertainty range is larger without the return current included (Case~1) than the one with it included (Case~2). This result indicates that including the return current provides tighter constraints on the overall star surface background temperature. Our maximum-likelihood analysis favors a magnetic inclination in which {the model without a return current yields $\iota \approx 90^{\circ}$, whereas the model including a return current gives a more plausible value of $\iota \sim 56^{\circ}$. This contrast strongly supports the necessity of incorporating return-current physics when using first-principles, physics-motivated hotspot models to interpret the observations.}
 Taken together, these findings suggest that the model with return current (Case~2) yields a more reasonable set of parameters, implying that return current effects may be essential for accurately modeling the observed emission.

Notably, the highest peaks for the $x$--, $y$--, and $z$-- shifts are all very close to zero, indicating that a minimal shift can explain the data quite well. It is also worth commenting on the inclination angle constraint. From gamma-ray observations \citep{Abdo_2013}, the magnetic inclination angle presented here is smaller other observation within 64\% uncertainty range. However, given that we are fixing the mass, radius, this is still remarkable and interesting.

In the top panel of Figure~\ref{fig:j0030_profile}, we present the hotspot configuration corresponding to the maximum-likelihood sample in Case~2, with the detailed model parameters listed in Case~2 column of Figure~\ref{fig3:posterior}. This configuration shows that both hotspots lie near the equatorial plane, and one appears {significantly} brighter {but smaller} than the other. The resulting structure is more complex than the configuration inferred by \citet{Vinciguerra2024}, which did not include the temperature gradient as incorporated in our approach. This increased complexity leads to a smaller magnetic inclination than the previous result, which is approximately $71.7^{\circ}$ inferred from \citet{Vinciguerra2024} ST+PDT maximum likelihood sample.%matching the peaks in the observer and magnetic inclination angles close to $\pi/2$. 

On the bottom of Figure \ref{fig:j0030_profile} compares the 1-D pulse profile data with the pulse profiles corresponding to the maximum likelihood samples in with and without return current cases. The bottom panel shows the residuals between the model and the data. The maximum-likelihood model explains the data well, with the main discrepancy occurring slightly in the valley and peak of the pulse profile. Given that we fixed the mass, radius, this is a good result achieved with a minimal number of parameters. There are still some degrees of freedom, such as the temperature normalization constant of each hotspot and the temperature-current relation polynomial index.

This minimal model demonstrates that Case 1 and Case 2 produce similar maximum-likelihood results with comparable residuals. Overall, the pulse profile of J0030 can potentially be understood within the framework of this off-center dipole field with a good fit, this result also match with the analysis in \citet{Vinciguerra2024}. A comprehensive analysis, including a detailed treatment of the joint analysis of the energy spectrum and the background, will be pursued in future studies.
\subsection{Inference results: PSR J0437--4715}
In the analysis of source J0437, We applied the same inference procedure as for J0030, with the exception of predefined model parameters—namely, mass $M_{\star}$, radius $R_{\star}$, Earth distance $D$, and coordinate frequency $\nu$—which differ from those used in the J0030 analysis. In this study, we keep consistent with the inference set-up for Earth distance and coordinate frequency as described in \citet{Choudhury_2024} and adopt their best-model result for mass and radius as our standard inputs, as detailed in Table \ref{tab1:xpsi_para}. Figure~\ref{fig3:posterior_j0437} displays the posterior distributions of the model parameters for the both cases: with and without return current. In both scenarios, the highest likelihood sample is highlighted in the table located at the top-right corner of the figure. The radio band observations provide a precise estimate of the observer inclination angle, approximately $ \sim 0.741\,\mathrm{rad}$, which is evident in the tighter constraint of this parameter when the return current is included. Although the inference is not yet fully converged, this preliminary result is encouraging.

For the magnetic inclination angle $\iota$, similar to the J0030 case, the posterior distribution for Case 2 exhibits a pronounced peak near $1.0$ {with a wide peak}, suggesting that this parameter remains less constrained by the current observations.   {The 2D posterior distribution features both the observer and magnetic inclinations peaking near $1.0\, \mathrm{rad}$, indicating a nontrivial configuration.} The posterior distributions for the dipole center shifts in the $x$--, $y$--, and $z$--directions are generally broader and the central value are larger compared to the J0030 inference result. This suggests that a larger dipole center shift is required to adequately explain the J0437 data. Furthermore, in direct comparisons between models with and without a return current, the inclusion of a return current {can help the dipole-shift posterior contract to a predominant peak} and simultaneously yields tighter constraints across the parameter space, {whereas, in the case without a return current, the posterior is far more dispersed and exhibits poor convergence, further highlighting the importance of including a return current. Collectively, these results indicate improved identifiability of the parameters when return-current physics is incorporated.} This aligns with our findings from the J0030 analysis, indicating a preference for the model with return current due to its more constrained parameter space and better alignment of the maximum-likelihood observer inclination value with radio observations. 

All dipole-shift parameters in both models are strongly decoupled from their priors and exhibit sharply peaked posteriors. For the surface temperature outside the hotspot, $T_s$, the posterior distributions are nearly identical across the two scenarios, {however, in the model without a return current the distribution is noisier, with some samples extending to very large $T_s$ and a small secondary peak near $T_s \sim 4$.}
However, for J0437, it is important to note that our inferred observer inclination is roughly consistent with the prior used in \citet{Choudhury_2024}, which is informed by radio observations \citep{Reardon_2024}. However, if we were to restrict the prior to maintain this consistency, our inference would fail to adequately explain the observational data. There are two possible explanations for this discrepancy: first, the temperature map reported in \cite{Choudhury_2024} includes a ring-shaped hotspot accompanying a circular hotspot, introducing distinct multipolar features that an off-center dipole model cannot fully capture. Second, \cite{Choudhury_2024} assumes that the radio-observed orbital inclination corresponds directly to the observer inclination angle, based on the assumption that the millisecond pulsar spin axis has aligned with the spin axis of its binary system over long-term evolution. Our results suggest a more flexible assumption, allowing for a potential misalignment.

The resulting hotspot configuration, derived from the highest-likelihood sample, is illustrated in the top of Figure~\ref{fig:shiftcombined_J0437} for the case that includes return current. The detailed parameter setup is also shown in the top-right corner of Figure~\ref{fig3:posterior_j0437} for case 2. This configuration indicates that {the northern hotspot exhibits a ring-shaped bright pattern, whereas the southern hotspot spans a larger area but is substantially fainter. Moreover, an additional asymmetry appears as a small emission region in the southeastern quadrant. This morphology is particularly intriguing, since the best-fitting result of \citet{Choudhury_2024} likewise favors a configuration that is qualitatively similar and features an asymmetric hotspot pattern. To better demonstrate how an off-centered dipole field yields the observed asymmetric hotspot geometry, we provide a companion video that visualizes the evolution of the best-fitting hotspot configuration as the off-center distance is increased to its optimal value. The video is available on Zenodo: \dataset[doi:10.5281/zenodo.16548609]{https://doi.org/10.5281/zenodo.16548609}.
} In addition, each hotspot exhibits an important arched component, a feature only arising from the inclusion of return current.

Comparing the light curves corresponding to the return-current and no-return-current cases against the observational data reveals that, although both scenarios can produce reasonably convergent posterior distributions, neglecting the return current leads to a noticeably poorer fit—particularly around the half-integer phase. In the no-return-current configuration, an deviation appears at this phase, resulting in larger residuals. This outcome underscores the importance of accounting for the return current in this computation.

In the return-current scenario, the maximum-likelihood sample continues to align reasonably well with the pulse profile data. Together with the posterior distribution in Figure~\ref{fig3:posterior_j0437}, these results suggest that an off-center dipole field configuration, when augmented by return-current effects, can capture the majority of the system’s X-ray emission features.

In all of our analyses, the high computational cost of these inferences has prevented being rigorously tested across multiple sampler configurations. The main objective of this work is to propose a new physics-motivated off-center-dipole field modeling procedure and demonstrate its feasibility using a state-of-the-art inference pipeline. {Consequently, at this proof‑of‑concept stage, the parameter estimates should be regarded as illustrative rather than definitive: the chains achieve a plausibly reasonable precision and display apparent convergence, yet a rigorous, multi sampling setup convergence demonstration is deferred to future work.} Future studies will undertake more advanced, large-scale inference runs that relax the currently fixed mass–radius inputs, in favor of treating the complete set of spacetime-related properties as free parameters to derive comprehensive inference once this physics-motivated framework is fully refined.

\section{Discussion and Conlusion}
\label{sec:discussion}

In this study, we employed an off-center dipole magnetic field model to investigate the temperature distribution on the surfaces of neutron stars, focusing on two key sources: PSR J0030+0451 and PSR J0437--4715. Our approach builds on recent development of the theoretical understanding of pulsar magnetospheres and surface heating due to magnetospheric current. Using the approximate force-free dipole solution, our model explicitly computes the temperature distribution of polar caps for an arbitrary off-center magnetic dipole moment of the neutron star, with or without the volumetric return current.
% builds on the modeling of X-ray pulse profiles, a technique crucial for constraining neutron star properties, including mass, radius, and hotspot temperature distributions. 
The off-center dipole model provides a simplified yet powerful framework for capturing the essential features of the magnetic field configuration, which influences the X-ray emission observed from these stars. 

For PSR J0030+0451, our results suggest that the inferred hotspot configuration within the context of an off-center dipole field can yield pulse profiles largely consistent with the observation data. The temperature gradient observed in our model introduces a {smaller} magnetic inclination than that inferred from 
% NICER 
Fermi gamma-ray
data~\citep{Abdo_2013, Chen_2020}, indicating that higher-order multipole moments might still be necessary to fully account for the star's multi-wavelength emission.

As for PSR J0437--4715, the inferred hotspot configuration is notably asymmetric, with one hotspot {exhibiting an interesting asymmetric structure that includes a small component located in the southeastern quadrant of the hotspot. This configuration surprisingly aligns with the best-fit hotspot configuration reported by \citet{Choudhury_2024}.} In this particular case, we see that the model with volumetric return current fits the data significantly better, indicating that it may be non-negligible at least in some scenarios.
% but suggests that an off-center dipole field model can capture the majority of this system's features. In this study , 

Our results demonstrate that the off-center dipole field model, despite its simplicity, is already capable of reproducing the observed pulse profiles of both sources with a considerable degree of accuracy. 
However, for both sources studied in this paper, there are features that are not fully accounted by this simple model. For example, our fitted magnetic inclination for J0030 is somewhat small, which is in tension with its gamma-ray pulse profile. Our fitted observer inclination for J0437 not be completely consistent with the very well constrained geometry inferred from the orbital motion. Both cases likely can be further improved with a more rigorous fitting procedure that takes into account the energy-dependent pulse profile. But these inconsistencies may also point to the need for multipole magnetic configurations as those suggested by ~\citet{Bilous_2019}, ~\citet{Chen_2020} and ~\citet{Kalapotharakos2021}. 

In this study, we directly employed the force-free model developed by \citet{Gralla2017} to compute the magnetospheric current in the off-center dipole scenario. This solution relies on an empirical formula for a crucial quantity $\Lambda$ that determines the magnitude of the current. The empirical formula was found by comparing with first-principles PIC simulations, and its validity in the off-center dipole regime has not been confirmed. Future work should carry out a quantitative comparison between the solution we constructed with global PIC simulations, which may significantly improve the quality of our model.
% promote the derivation from the centered dipole scenario 

% in \cite{Gralla2017} to shifted-dipole regime, while the empirical relation that employed in the centered dipole field like \ref{lambda} is approximated from the first principle PIC simulation, the validiablity of this formula when promote to the off-center dipole field need further exploration with detailed PIC simulation, however, since judging from the multipole expansion form of off-center dipole \cite{P_tri_2016}, off-center dipole dominantly to be centered dipole components, and all the higher multi-pole components will decay to infinitesimal with increasing the distance from the star center, so with this argument, we argue this direct promotion of this formula to a shifted dipole case without change the form is reasonable but need further clarification.

%We did not include hotspot temperature normalization as a parameter. Future work should include it, and also include the X-ray spectrum as part of the data fitting process, as it can help constrain the hotspot temperature.
% In the current study, 
Similarly, we have employed a non-relativistic definition of the Euler potential $\alpha$ to account for the modifications introduced by the shifted dipole. This assumption ignores the general relativistic correction to the dipole solution, which is difficult to incorporate as the dipole center no longer coincides with the mass center. Nevertheless, we have compared the GR dipole solution from~\citet{Gralla2017} with our non-relativistic model, and the deviation is up to around $3\%$. They do not significantly alter the overall distribution of the temperature map or the area of the bright regions. Therefore, we argue that even in the absence of a comprehensive treatment of GR effects, our theoretical computations remain reliable and approximate the hotspot geometry adequately. A more comprehensive GR solution could be pursued in future studies.
% This assumption aligns with the approach adopted in real-case studies. Although the complete general relativity (GR) definition of $\alpha$ could be intricate and beyond the scope of our work, previous studies \cite{P_tri_2016} have explored a multipolar expansion of the off-center dipole. However, a unified analytic expression of the off-center dipole field under GR remains still not clear.

% Nevertheless, we contend that the incorporation of GR effects into the definition of $\alpha$, as indicated by the factor correlated to $f(M,r)$ in equation \ref{5} compared to equation \ref{7}, is limited in scope. Upon reintroducing this factor into the definition and substituting gravitational mass and radius values before the shift (as the gravitational center should remain unchanged), the modifications are minimal (with order of 3\%). 
Another crucial assumption in this paper is the simple surface heating model where the power is directly proportional to the electric current $j$. Since the polar cap heating is ultimately controlled by the complex QED pair cascade process, it is natural to expect that the temperature profile has more complicated dependence on $j$ than Equation~\eqref{eq:T-scaling}. To the best knowledge of the authors, such a phenomenological model has yet to be developed. A detailed microphysical description of how the Poynting flux is dissipated and converted to plasma kinetic energy, which in turn heats the surface of the neutron star, may significantly improve the reliability of future physics-based X-ray pulse profile modeling.

% \AC{[Rewrite this paragraph to discuss how the heating model can determine the temperature distribution]} In the current study, no normalization constant that introduced in \cite{Lockhart2019} has been introduced in the inference, while this could be crucial since the specific heating proportional coefficient is still under-determined, even for same neutron star different hotspots may have different normalization constants to reflect the different heating and atomspherical effect. However, in a real inference, releasing this normalization as free parameters would enable exploration of the correlated external magnetosphere properties by incorporating them into certain heating and atmospheric models. To avoid complexity, the details of this complication were not explored in this study, and they were simply replaced with two free parameters.

%We did not use a fully general relativistic expression for alpha and beta. Since GR effects are already partly taken into account in the expressions for $j$, we have included the most important effects. The deviation of GR version of alpha from our exression is on the order of 3\%.

%Better theoretical models of the pulsar magnetosphere can also improve our model. In particular, the heating model we adopt in this paper is rather crude. A better model that can estimate the realistic heating power as a function of $j/c\rho_\mathrm{GJ}$ and local field line curvature $\rho$ can significantly improve the surface temperature distribution.

At this stage, conducting a full-scope inference using the energy-resolved 2D pulse profile data with the physics-motivated hotspot model from this paper remains a challenging task due to the extreme computational cost associated with the large dimension of data and computation requirement of the temperature map. Further optimization of the pipeline that computes the energy-resolved pulse profiles from the physical parameters is needed. We defer this nontrivial problem to a future study.

\section{Acknowledgements}
The authors would like to thanks for the insightful discussions with Anna Watts and Sam Gralla. {We also thank the anonymous referee for their comments that helped improve the manuscript}. Special thanks to Pushpita Das, Tuomo Salmi, and Bas Dorsman for their assistance with the usage of the \textit{X-PSI} software. Additionally, thanks to Serena Vinciguerra for generously providing inference data and guidance on implementing geometric computations. This research is supported by NSF grants DMS-2235457 and AST-2308111, and NASA grant 80NSSC24K1095.


\bibliographystyle{aasjournal}
\bibliography{main}{}



\end{document}